\documentclass[a4paper,11.5pt]{article}
\pdfoutput=1
\usepackage{graphicx}
\usepackage{epstopdf}
\usepackage[labelfont=bf]{caption}
\usepackage{amssymb,amsmath,fixmath}
\usepackage[numbers,compress]{natbib}
\usepackage{multirow}
\usepackage{booktabs}
\usepackage{adjustbox}

\title{Regular and chaotic orbits in the dynamics of exoplanets}
\author{Kyriaki I. Antoniadou\thanks{\emph{E-mail:} kyant@auth.gr} \\
\small{Section of Astrophysics, Astronomy and Mechanics}, \\ \small{Department of Physics, Aristotle University of Thessaloniki,} \\
\small{Thessaloniki, 54124, Greece}\\}                     
\date{\vspace{-5ex}}
\begin{document}

\maketitle

\begin{center}
`The final publication is available at Springer via http://dx.doi.org/10.1140/epjst/e2016-02651-6'
\end{center}
\begin{abstract}
Many of exoplanetary systems consist of more than one planet and the study of planetary orbits with respect to their long-term stability is very interesting. Furthermore, many exoplanets seem to be locked in a mean-motion resonance (MMR), which offers a phase protection mechanism, so that, even highly eccentric planets can avoid close encounters. However, the present estimation of their initial conditions, which may change significantly after obtaining additional observational data in the future, locate most of the systems in chaotic regions and consequently, they are destabilized. Hence, dynamical analysis is imperative for the derivation of proper planetary orbital elements. We utilize the model of spatial general three body problem, in order to simulate such resonant systems through the computation of families periodic orbits. In this way, we can figure out regions in phase space, where the planets in resonances should be ideally hosted in favour of long-term stability and therefore, survival. In this review, we summarize our methodology and showcase the fact that stable resonant planetary systems evolve being exactly centered at stable periodic orbits. We apply this process to co-orbital motion and systems HD 82943, HD 73526, HD 128311, HD 60532, HD 45364 and HD 108874.
\end{abstract} 


\section{Introduction}\label{intro}
Nowadays, the quest of exoplanets plays a key role in the field of astronomy. Heretofore (December 2015), within 1293 planetary systems that have been discovered, 504 out of which being multiple planet systems, 2041 extrasolar planets are confirmed. The most important attributes of them, forcing scientific community to think of our Solar System as an exception rather than a typical planetary system, are: the orbital eccentricities, the mass distribution and the semi-major axes. 

We observe that a lot of planets have masses very similar to those of planets in our Solar System, however, they have quite eccentric orbits and they are located closer to their host stars. The majority has a Jovian mass value, an orbital period of 8-10 days and circular orbits. Conversely, a minority of them is evolving on highly eccentric and/or inclined orbits. What is more, multiple planet systems seem to be locked in MMR with the majority of which in 2/1 and by descending order in 3/2, 5/2, 3/1, 4/1 and 4/3 \cite{lask09,corr10,rein10,lovis11,camp13}. However, depending on the method of detection, the observational errors can usually lead to large uncertainties of the published orbital elements and planetary masses \cite{wit13}.  This lack in precise knowledge of parameters in turn, in most cases, locates the exoplanets in chaotic regions, where their survival is not guaranteed. 

Observational data for Kepler Objects of Interest (KOIs) indicating trappings in 1/1 MMR of stable and mutually inclined planets are mostly rejected by astronomers as false-positive, although they are theoretically predicted \cite{avv14}. A very recent analysis \cite{lillo11} confirmed those predictions and requested more dynamical studies. 

Regarding the formation of inclined exoplanets, some possible mechanisms that lead to excitation of planetary inclinations are the planetary scattering \cite{mawei02,chaford08}, the differential migration \cite{thommes03,leetho09} and the tidal evolution \cite{cor11}. In these processes, resonant capture can occur and the evolution of the exoplanetary system can be associated with particular families of either co-planar or even, mutually inclined periodic orbits \cite{mebeaumich03,lee04,bmfm06,hadjvoy10,hv11,vat14}.

Therefore, extrasolar planetary systems must undoubtedly be tested in reference to their long-term dynamical stability, the deviations to their orbital elements should be revised (or proposed, in case the observational method fails to provide them) in order for such distributions of planets to be justified and trappings, in MMR being observed, to be achieved.

It is well-known, that planets evolving in MMR prompt the investigation of resonant dynamics in the framework of the general three-body problem (GTBP) \cite{mbf06,vkh09,delaco12,av12,av13,delaco14}. Utilizing the GTBP as a model, the computation of families of periodic orbits in a suitable rotating frame of reference can help ascertain information regarding the phase space in their vicinity \cite{numan2014}.

In this review, we firstly provide the fundamentals of our dynamical analysis, apply it to co-orbital motion and exoplanetary systems HD 82943, HD 73456, HD 128311 (locked in 2/1 MMR), HD 60532 (trapped to 3/1 MMR), HD 45364 (evolving in 3/2 MMR) and HD 108874 (captured in 4/1 MMR) and exhibit the fact that families of stable periodic orbits consist the backbone of stability domains in phase space; the essential regions where planetary systems can evolve regularly for long timescales.

\section{Model, periodic orbits and stability}\label{sec1}
We herein introduce the model used to simulate three body systems and present the way orbits of long-term stability are determined, after having computed the families of periodic orbits and determined their linear stability.  

\subsection{The spatial general three body problem}\label{sec2}
Let us consider two mutually inclined planets revolving around a star under their mutual gravitational attraction and consider their masses, $m_0$, $m_1$ and $m_2$ as point masses. Subscripts 0, 1 and 2 shall always refer to the star, $S$, the inner planet, $P_1$ and the outer one, $P_2$, respectively. We should note though, that chaotic motion may change the initial location of the planets. These bodies move in an inertial frame of reference, $OXYZ$, whose origin is their fixed center of mass and its $Z$-axis is parallel to the constant angular momentum vector of the system. Given the former, we can determine the position and velocity of the star via those of the planets. Consequently, the system will be of 6 degrees of freedom and possess the integrals of the energy and the angular momentum.

The degrees of freedom can be reduced to 4 by introducing a suitable rotating frame of reference, $Gxyz$, such that: a) its origin is the center of mass of $S$ and $P_1$, b) these bodies shall always move on the $xz$-plane and c) its $Gz$ is parallel to the $OZ$. Also, by defining the rotation of this frame through the angle $\theta$ between $OX$ and $Gx$ axes and by assuming always that $\theta(0)=0$, it always holds that $y_1=0$. Thus, the Lagrangian that corresponds to the system, having previously normalized the gravitational constant $\mathcal G$ and the total mass, $m=m_0+m_1+m_2$ to unity, is:
\begin{equation}
\begin{array}{l}
L=\frac{\displaystyle 1}{\displaystyle 2} \mu[a(\dot x_1^2+\dot z_1^2+x_1^2\dot \theta^2)+\displaystyle b [(\dot x_2^2+\dot y_2^2+\dot z_2^2)+\dot\theta^2(x_2^2+y_2^2)+2\dot\theta(x_2\dot y_2-\dot x_2y_2)]]-V,
\label{Lagrangian}
\end{array}
\end{equation}
where
\begin{equation}
\begin{array}{llll}
V=-\frac{\displaystyle m_0 m_1}{\displaystyle r_{01}}-\frac{\displaystyle m_0 m_2}{\displaystyle r_{02}}-\frac{\displaystyle m_1 m_2}{\displaystyle r_{12}},&a=m_1/m_0,&b=m_2/m,&\mu=m_0 + m_1
\end{array}
\end{equation}
with $r_{01}^2=(1+a)^2(x_1^2+z_1^2)$, $r_{02}^2=(a x_1+x_2)^2+y_2^2+(a z_1+z_2)^2$ and $r_{12}^2=( x_1-x_2)^2+y_2^2+ (z_1-z_2)^2$.

Since $\theta$ is an ignorable variable, the system bears the following angular momentum integral
\begin{equation}
p_\theta=L_Z=\mu[ax_1^2\dot \theta+\displaystyle b[\dot \theta (x_2^2+y_2^2)+(x_2\dot y_2-\dot x_2y_2)]]=\textnormal{const.}
\label{Lz}
\end{equation}
and since no angular momentum is considered regarding the $Gx$ and $Gy$ axes ($L_X=L_Y=0$) two more restrictions arise:
 \begin{equation}
\begin{array}{l}
 z_1=\frac{b}{a} \left ( \frac{y_2\dot z_2-\dot y_2 z_2}{x_1 \dot \theta}-\frac{x_2z_2}{x_1} \right ), \\
\dot z_1=\frac{b}{a x_1}(\dot x_2 z_2-x_2\dot z_2 -\dot{\theta}y_2z_2)+\frac{\dot x_1}{x_1} z_1. 
\label{Lxy}
\end{array}
\end{equation}
Thus, such a spatial three body system can be simulated by 4 equations of motion
\begin{equation}
\begin{array}{l}
\ddot x_1=-\frac{m_0 m_2 (x_1-x_2)}{\mu r_{12}^3}-\frac{m_0 m_2 ( a x_1+x_2)}{\mu r_{02}^3}-\frac{\mu x_1}{r_{01}^3}+x_1\dot{\theta}^2\\ 
\ddot x_2=\frac{m m_1 (x_1-x_2)}{\mu r_{12}^3}-\frac{m m_0 (a x_1+x_2)}{\mu r_{02}^3}+x_2\dot{\theta}^2+2 \dot y_2 \dot{\theta}+y_2\ddot{\theta}\\
\ddot y_2=-\frac{m m_1 y_2}{\mu r_{12}^3}-\frac{m m_0 y_2}{\mu r_{02}^3}+y_2\dot{\theta}^2-2 \dot x_2 \dot{\theta}-x_2\ddot{\theta}\\
\ddot z_2=\frac{m m_1 (z_1-z_2)}{\mu r_{12}^3}-\frac{m m_0 (a z_1+z_2)}{\mu r_{02}^3}
\label{eq}
\end{array}
\end{equation}

The planar problem is simply derived by setting the third dimensions ($z_{1,2}$ and $\dot z_{1,2}$) equal to zero. Nevertheless, we should remark some essential differences between the restricted\footnote{One body has a negligible mass, in comparison to the rest, the \textit{primaries}, in a way that does not perturb their motion, which can be either circular (CRTBP) (under a constant radius and $\dot \theta=1$) or elliptic (ERTBP).} (3D-RTBP) and the general spatial three body problem described in the rotating frame. On the GTBP the primaries do not remain fixed on the rotating $Gx$-axis, but move on the rotating plane $xz$. The rotating frame does not revolve on a constant angular velocity since 
\begin{equation}
\dot \theta=\frac{\frac{p_\theta}{\mu}-b (x_2\dot y_2-\dot x_2 y_2)}{a x_1^2+ b (x_2^2+y_2^2)}
\label{dth}
\end{equation}
and its origin $G$ does not remain fixed with respect to $OXYZ$.

Last but not least, it has been proved  \cite{moi} that the equations of motion (both in the planar and the spatial case) remain invariant under a change in units, if for instance, one wishes to use different units of time, ($T^*$), masses, $m^*$,  distances, $a^*$, etc. Particularly, the term $\frac{T^{*2} {\mathcal {G}} m^*}{a^{*3}}$ should equal to unity.

\subsection{Periodic orbits and mean-motion resonances}\label{sec3}
Given a Poincar\'e surface of section in phase space, e.g. $\hat\pi=\{y_2=0,\dot y_2>0\}$, considered in $Gxyz$, the \textit{periodic orbits} are defined as the fixed or periodic points on this map, as long as they fulfill the periodicity conditions 
\begin{equation} \label{apocon}
\begin{array}{l}
x_1(0)=x_1(T),\; x_2(0)=x_2(T), \; z_2(0)=z_2(T),\\
\dot x_1(0)=\dot x_1(T),\; \dot x_2(0)=\dot x_2(T), \; \dot z_2(0)=\dot z_2(T), \dot y_2(0)=\dot y_2(T),
\end{array}
\end{equation}
provided that $y_2(0)=y_2(T)=0$ and $T$ is the period. 

The Lagrangian \eqref{Lagrangian} is invariant under four transformations and therefore, there exist four \textit{symmetries} to which the periodic orbits obey:
\begin{equation} \label{symme}
\begin{array}{ll}
\Sigma_1:(x_1,x_2,y_2,z_2,t)\rightarrow(x_1,x_2,-y_2,z_2,-t),& \Sigma_2:(x_1,x_2,y_2,z_2,t)\rightarrow(x_1,x_2,-y_2,-z_2,-t)\\
\Sigma_3:(x_1,x_2,y_2,z_2,t)\rightarrow(-x_1,-x_2,y_2,z_2,-t),& \Sigma_4:(x_1,x_2,y_2,z_2,t)\rightarrow(-x_1,-x_2,y_2,-z_2,-t).
\end{array}
\end{equation} 

\begin{figure}
\begin{center}
$\begin{array}{ccc}
\includegraphics[width=5.25cm,height=5.25cm]{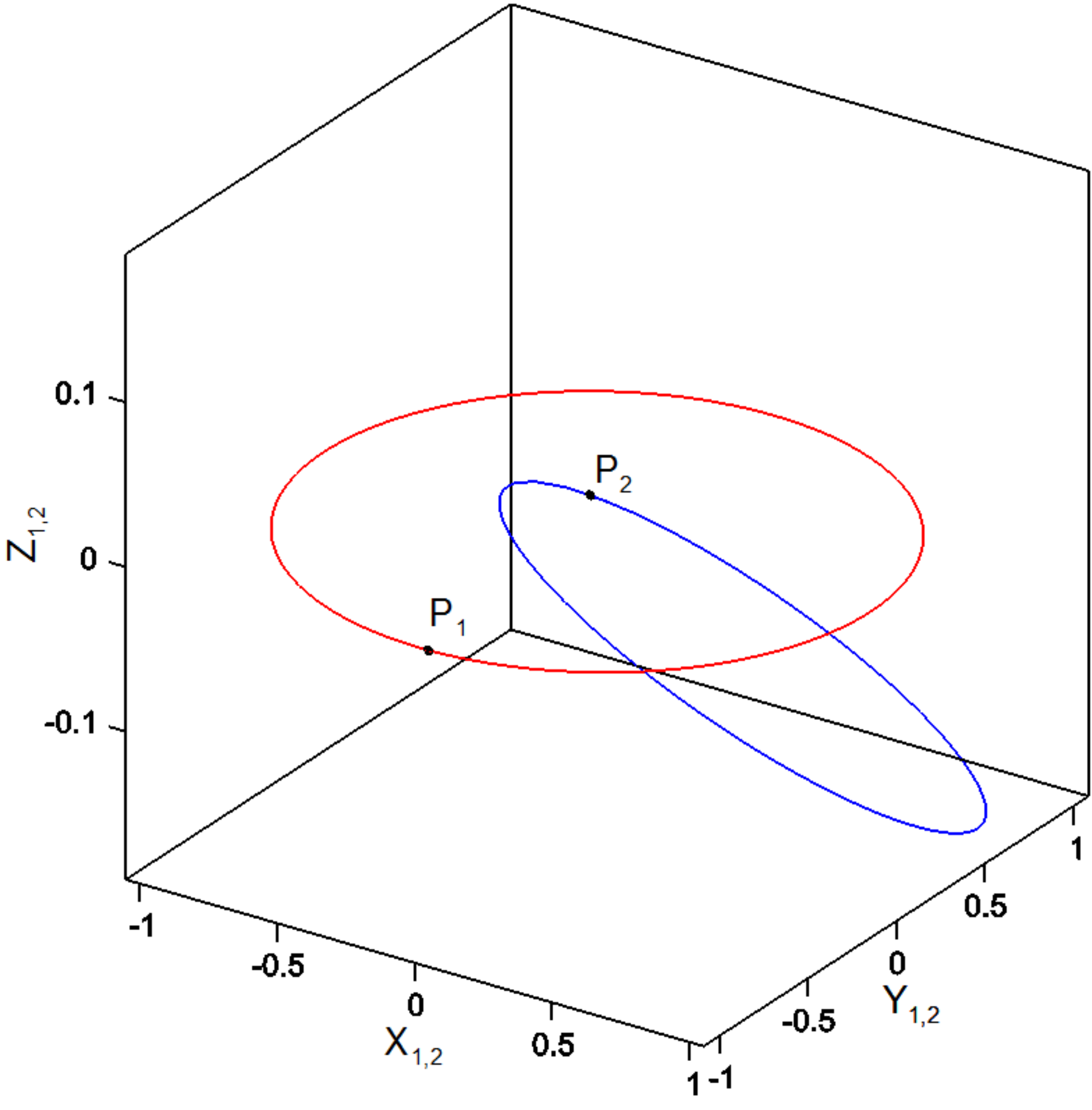}  &\;&
\includegraphics[width=5.5cm,height=5.5cm]{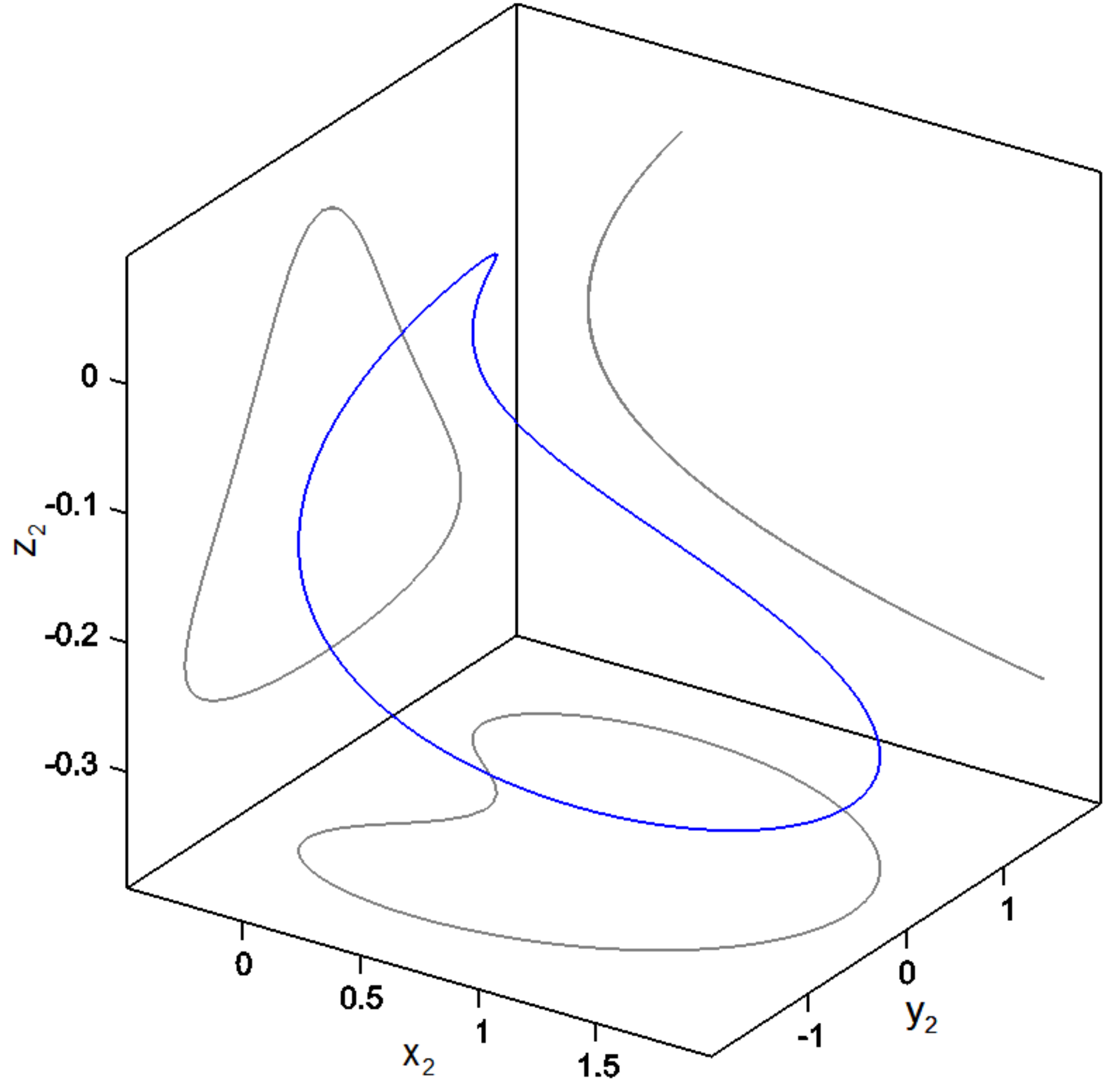}  \\
\textnormal{(a)}  & \;&\textnormal{(b)} 
\end{array} $
\end{center}
\caption{A $xz$-symmetric $1/1$ periodic orbit of planetary mass ratio $\rho=m_2/m_1=0.01$ and orbital elements:
$a_1=1.00, e_1=0.056, i_1=0.061^\circ, \varpi_1=90^\circ, \Omega_1=90^\circ, M_1=180^\circ, a_2=1.00, e_2=0.782, i_2=9.87^\circ, \varpi_2=90^\circ, \Omega_2=270^\circ, M_2=0^\circ$.
The planetary orbits in {\bf a} the inertial and {\bf b} the rotating frame together with the projection to the planes of the frame.}
\label{inrot}
\end{figure}

If a periodic orbit is invariant under $\Sigma_1$ is called \textit{xz-symmetric} (see Fig. \ref{inrot}) and is represented by the initial conditions
\begin{equation}
\begin{array}{llll}
x_1(0)=x_{10}, &\quad x_2(0)=x_{20}, &\quad y_2(0)=0 & \quad z_2(0)=z_{20},  \\
\dot x_1(0)=0, &\quad \dot x_2(0)=0, &\quad \dot y_2(0)=\dot y_{20}, &\quad \dot z_2(0)=0.
\label{xzsym}
\end{array}
\end{equation}
Accordingly, if it is invariant under $\Sigma_2$ is called \textit{x-symmetric} and is represented by the set 
\begin{equation}
\begin{array}{llll}
x_1(0)=x_{10}, &\quad x_2(0)=x_{20}, &\quad y_2(0)=0, &\quad z_2(0)=0,\\
\dot x_1(0)=0, &\quad \dot x_2(0)=0, &\quad \dot y_2(0)=\dot y_{20}, &\quad \dot z_2(0)=\dot z_{20}.
\label{xsym}
\end{array}
\end{equation}

Similarly, we can define \textit{yz-} or \textit{y-symmetric} periodic orbits.

All the rest initial conditions \eqref{apocon} correspond to \textit{asymmetric} periodic orbits.

By changing the value of $z_2$ (for the $xz$-symmetry) or $\dot z_2$ (for the $x$-symmetry) a \textit{monoparametric family} of periodic orbits is formed. Also, a  monoparametric family can be formed by changing the mass of a planet, $P_1$ or $P_2$, but keeping the value $z_2$ (or $\dot z_2$) constant. 

In the 3D-CRTBP, $x_1$ is constant defined by the normalization adopted for the system and in our computations is taken equal to $1-\frac{m_1}{m_0+m_1}$. Thus, the $xz$-symmetric periodic orbits can be represented by a point in the three-dimensional space of initial conditions
$\Pi_3=\{(x_{20},z_{20},\dot{y}_{20})\}$ and the $x$-symmetric ones by $\Pi'_3=\{(x_{20},\dot{y}_{20},\dot{z}_{20})\}.$

Families of periodic orbits can either be \textit{circular} or \textit{elliptic}. The circular ones consist of symmetric periodic orbits and $\frac{n_1}{n_2}$ varies along them. The elliptic ones are formed by symmetric or asymmetric periodic orbits and along them the MMR $\frac{n_1}{n_2}=\frac{p+q}{p}=\rm{rational}$\footnote{If the periodic orbits correspond to planetary motion the mean-motion ratio is approximately equal to this rational number.}, q being the order of the MMR, remains almost constant. Therefore, it is the resonant periodic orbit, which identifies the location of the exact MMR in phase space. 

In the neighbourhood of the stable (see Sect. \ref{sec5}) resonant periodic orbits the resonant angles 
\begin{equation}
\begin{array}{l}
\theta _{1}=(p+q)\lambda _{2}-p\lambda _{1}-q\varpi _{1}\\ 
\theta _{2}=(p+q)\lambda _{2}-p\lambda _{1}-q\varpi _{2}\\
\theta _{3}=(p+q)\lambda _{2}-p\lambda _{1}-\tfrac{q}{2}(\varpi _{1}+\varpi _{2})
\end{array}
\end{equation}
and the apsidal difference, $\Delta\varpi$, librate about $0$ or $\pi$, if the orbit is symmetric, or around other angles, if it is asymmetric. A significant attribute of the latter ones, is the fact that they come in pairs; one being the mirror image of the other in phase space \cite{voyatzis08,avk11}. Thus, a precession of $2\pi-\varpi_{1,2}$ will also be observed and the location of mean anomalies in phase space at $2\pi-M_{1,2}$ will be valid, as well. 

The libration of the above resonant angles showcases the \textit{eccentricity resonance}(e-resonance).

There exist four different symmetric configurations, if we assume aligned, $\Delta\varpi=0$, and anti-aligned, $\Delta\varpi=\pi$ planets, which do not change along the families of periodic orbits. Thus, we represent them on the eccentricities plane. When $q=2k+1, k\in\mathbb{Z}^*$, we use the pair $(\theta_1,\theta_2)$ and when $q=2k, k\in\mathbb{Z}^*$, we use the pair $(\theta_3,\theta_1)$, in order to distinguish the different groups of families (see \cite{av13}). For small planetary masses, it has been shown by \cite{beau03,fer06} that the characteristic curves belonging to the same configuration differ one another in planetary mass ratio $\rho=\frac{m_2}{m_1}$.

When the two planets are not co-planar, we may introduce the resonant angles that define the \textit{inclination resonance} (i-resonance) for at least second order resonances  
\begin{equation}
\begin{array}{l}
\varphi _{11}=(p+q)\lambda _{1}-p\lambda _{2}-q\Omega _{1}\\ 
\varphi _{22}=(p+q)\lambda _{1}-p\lambda _{2}-q\Omega _{2}\\
\end{array}
\end{equation}
We may further define the mixed resonance angle
\begin{equation}
\begin{array}{l}
\varphi _{12}=(p+q)\lambda _{1}-p\lambda _{2}-\Omega _{1}-\Omega_2=(\varphi_{11}+\varphi_{22})/q
\end{array}
\end{equation}
as well as the zeroth order secular resonance angle
\begin{equation}
\begin{array}{l}
\varphi _{\Omega}=\Omega _{1}-\Omega_2=(\varphi_{11}-\varphi_{22})/q.
\end{array}
\end{equation}

\subsection{Order and chaos}\label{sec4}
Deciphering an 8D phase space, in the 3D-GTBP, in terms of location of regular or chaotic orbits can be demanding, unless a clue is given. Periodic orbits can guide this research and expedite the tracing of regions of stability. Evolution of dynamical systems in the neighbourhood of stable periodic orbits takes place within invariant tori and the orbits are quasi-periodic. In the vicinity of unstable periodic orbits, chaotic domains exist. Therefore, the characterization of the periodic orbits in this respect is imperative. We hereby define the linear horizontal and vertical stability and then, explore the extent of regular regions in phase space by the construction of dynamical stability maps.

\subsubsection{Linear horizontal stability}\label{sec5}

The linear stability of the spatial periodic orbits can be found by computing the 4 pairs of conjugate eigenvalues of the $8\times 8$ monodromy matrix of the variational equations of system \eqref{eq}. More precisely, if the variational equations and their solutions are of the form
\begin{equation}
\begin{array}{ccc}
\dot{\mathbold{\eta}}=\textbf{J}(t)\mathbold{\eta} & \Rightarrow & \mathbold{\eta}=\Delta(t)\mathbold{\eta}_0 
\end{array}
\end{equation}
where \textbf{J} is the Jacobian of the right part of the system's equations and $\Delta (t)$ the fundamental matrix of solutions, then, according to Floquet's theory, the deviations $\mathbold{\eta}(t)$ remain bounded, iff all eigenvalues (one pair is always equal to unity, due to the existence of the energy integral) lie on the unit circle \cite{marchal90,hadjbook06}. Then, the respective spatial periodic orbit is called linearly \textit{stable}. Otherwise, it is linearly \textit{unstable}.

The linear horizontal stability of the planar periodic orbits can be retrieved in the above mentioned way, if the third dimensions are set equal to zero. Then, the location of the 3 pairs of eigenvalues of the $6\times6$ monodromy matrix of the respective variational equations can classify the periodic orbits as linearly \textit{horizontally stable},  or linearly \textit{horizontally unstable}.

\subsubsection{Linear vertical stability}\label{sec6}

The linear vertical stability of a planar periodic orbit is computed, if we linearize the last equation of the set \eqref{eq} and compute the variational equations
\begin{equation}
\dot \zeta_1=\zeta_2 \ \ \ \dot \zeta_2=A \zeta_1 +B \dot \zeta_2
\label{z22}
\end{equation}
where
\begin{equation}
\begin{array}{l}
A=-\frac{m m_0 [1-\frac{m_2 ( \dot \theta x_2+\dot y_2)}{m \dot \theta x_1}]}{(m_0+m_1)[(\frac{m_1 x_1}{m_0}+x_2)^2+y_2^2]^{3/2}} -\frac{m m_1 [1+\frac{m_0 m_2 (\dot \theta x_2+\dot y_2)}{m m_1 \dot \theta x_1}]}{(m_0+m_1)[(x_1 - x_2)^2+y_2^2]^{3/2}}\\[0.3cm]
B=\frac{m_0 m_2 y_2}{(m_0 + m_1)\dot \theta x_1 [(x_1-x_2)^2+y_2^2]^{3/2}}-\frac{m_0 m_2 y_2}{(m_0 + m_1)\dot \theta x_1 [(\frac{m_1 x_1}{m_0}+x_2)^2+y_2^2]^{3/2}}.
\label{AB}
\end{array}
\end{equation}

Then, after obtaining the monodromy matrix, $\Delta(T)=\zeta_{ij},\  i,j=1,2$ we can compute either the pair of eigenvalues, or the \textit{vertical stability index}
\begin{equation}
a_v=\frac{1}{2}(\zeta_{11}+\zeta_{22}).
\label{av222}
\end{equation}

Any periodic orbit of the planar problems with $|a_v|=1$  is  {\em vertical critical} (v.c.o.) and can be used as a starting orbit for  analytical continuation to the spatial problems. If $|a_v|<1$ or $|a_v|>1$ the planar periodic orbit is characterized linearly \textit{vertically stable} or \textit{vertically unstable}, respectively.

\subsubsection{Dynamical stability maps}\label{sec7}

Dynamical stability maps (DS-maps) offer a straightforward visualization of phase space. Grids of initial conditions on a plane are created by varying two parameters of the system and keeping the rest variables fixed. Then, for each point of the grid a chaotic indicator is computed and according to the output the point is coloured. 

\begin{figure}
\begin{center}
$\begin{array}{ccc}
\includegraphics[width=5.5cm,height=5.5cm]{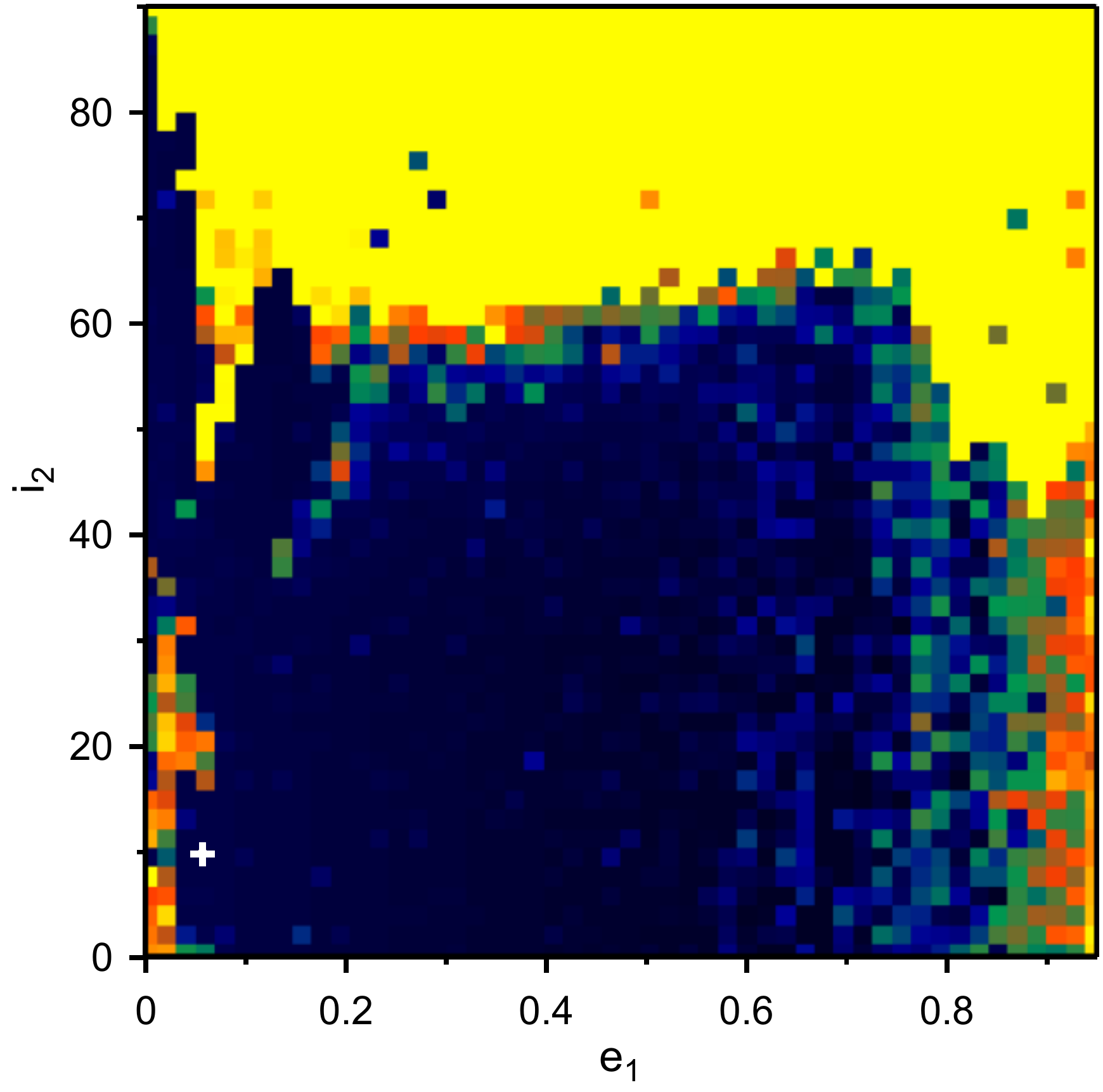}  &\;&\includegraphics[width=5.5cm,height=5.5cm]{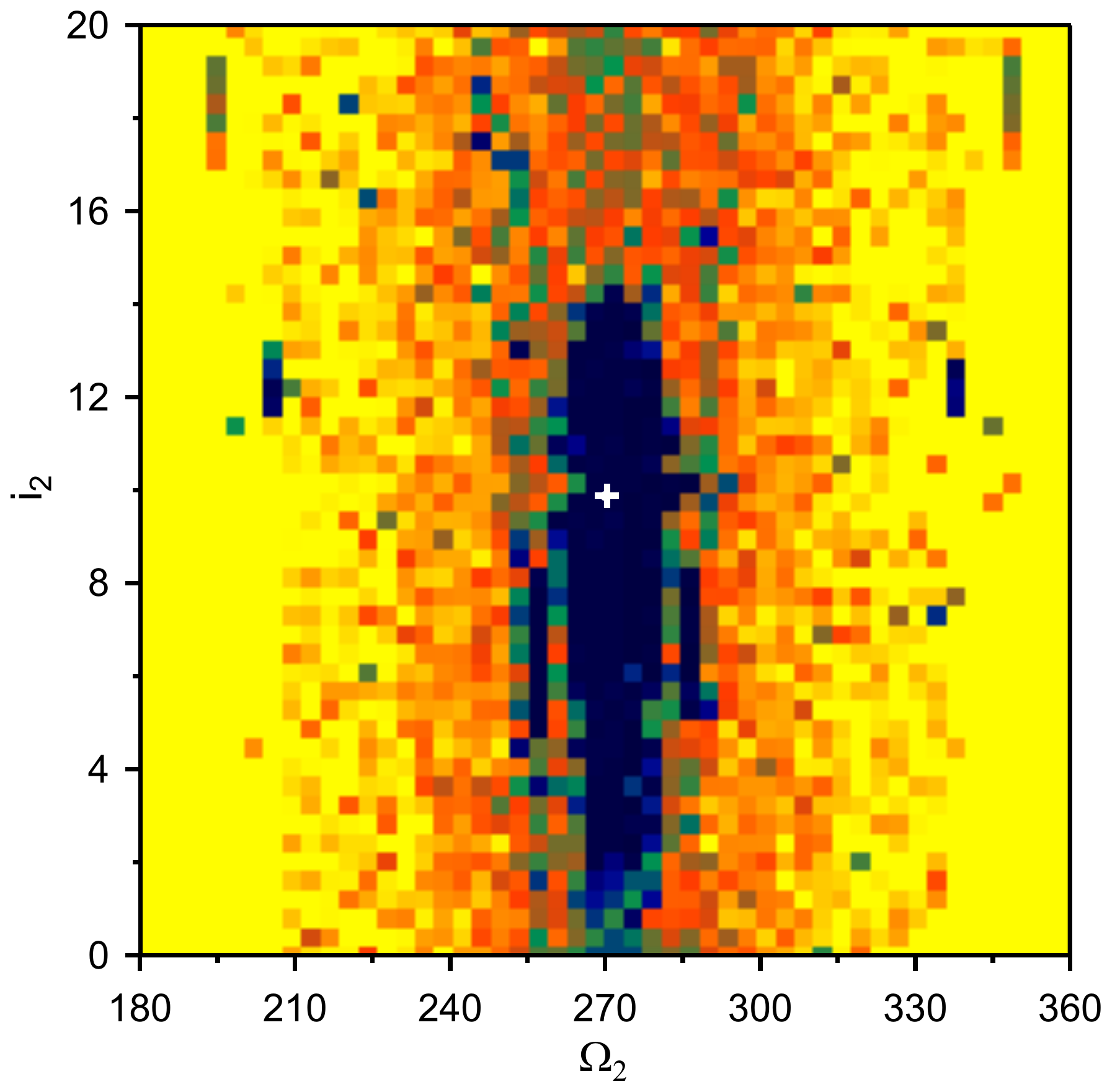}  \\
\textnormal{(a)}  & \;&\textnormal{(b)} \\
\includegraphics[width=5.5cm,height=5.5cm]{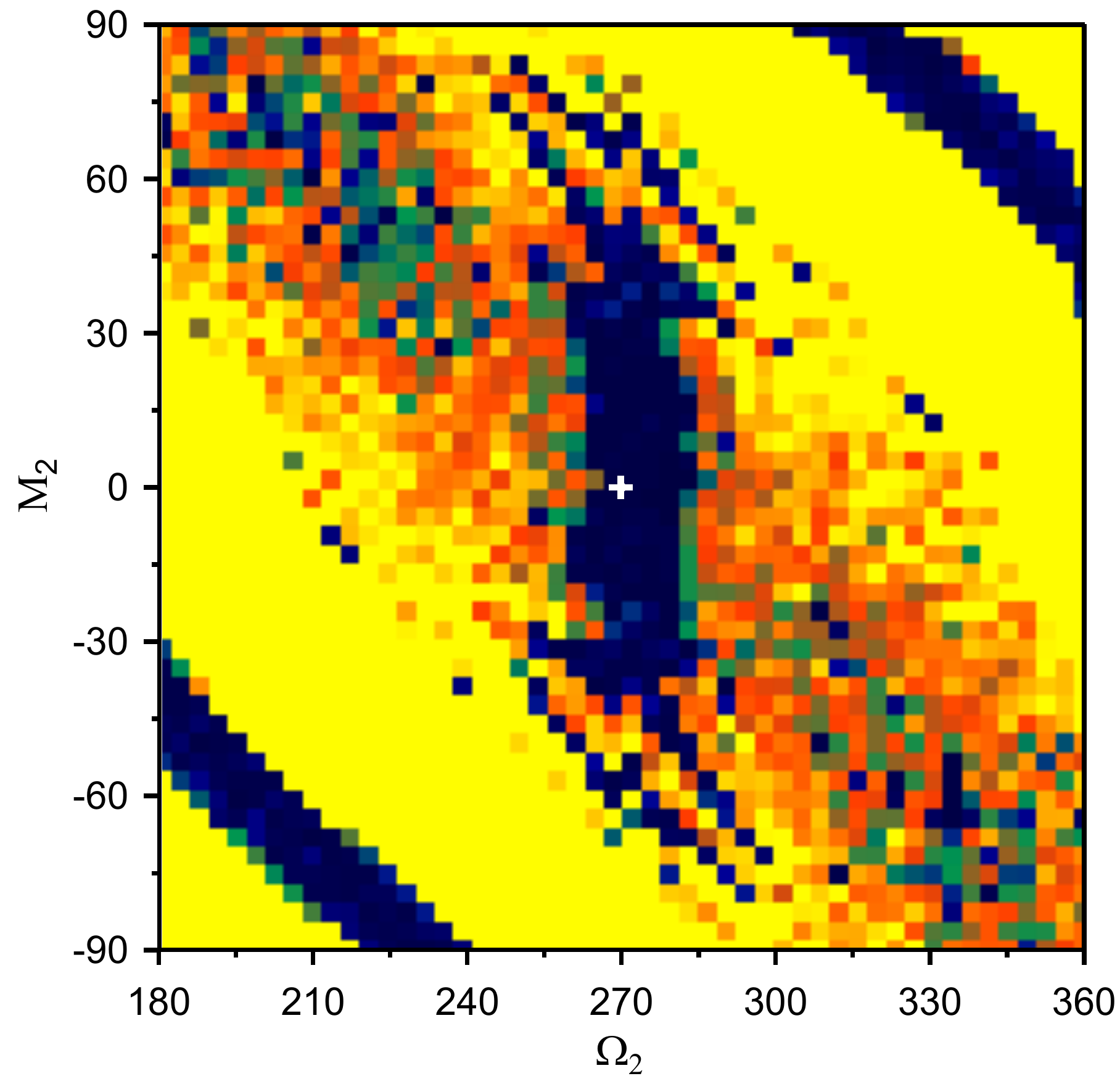} &\;& \includegraphics[width=5.5cm,height=5.5cm]{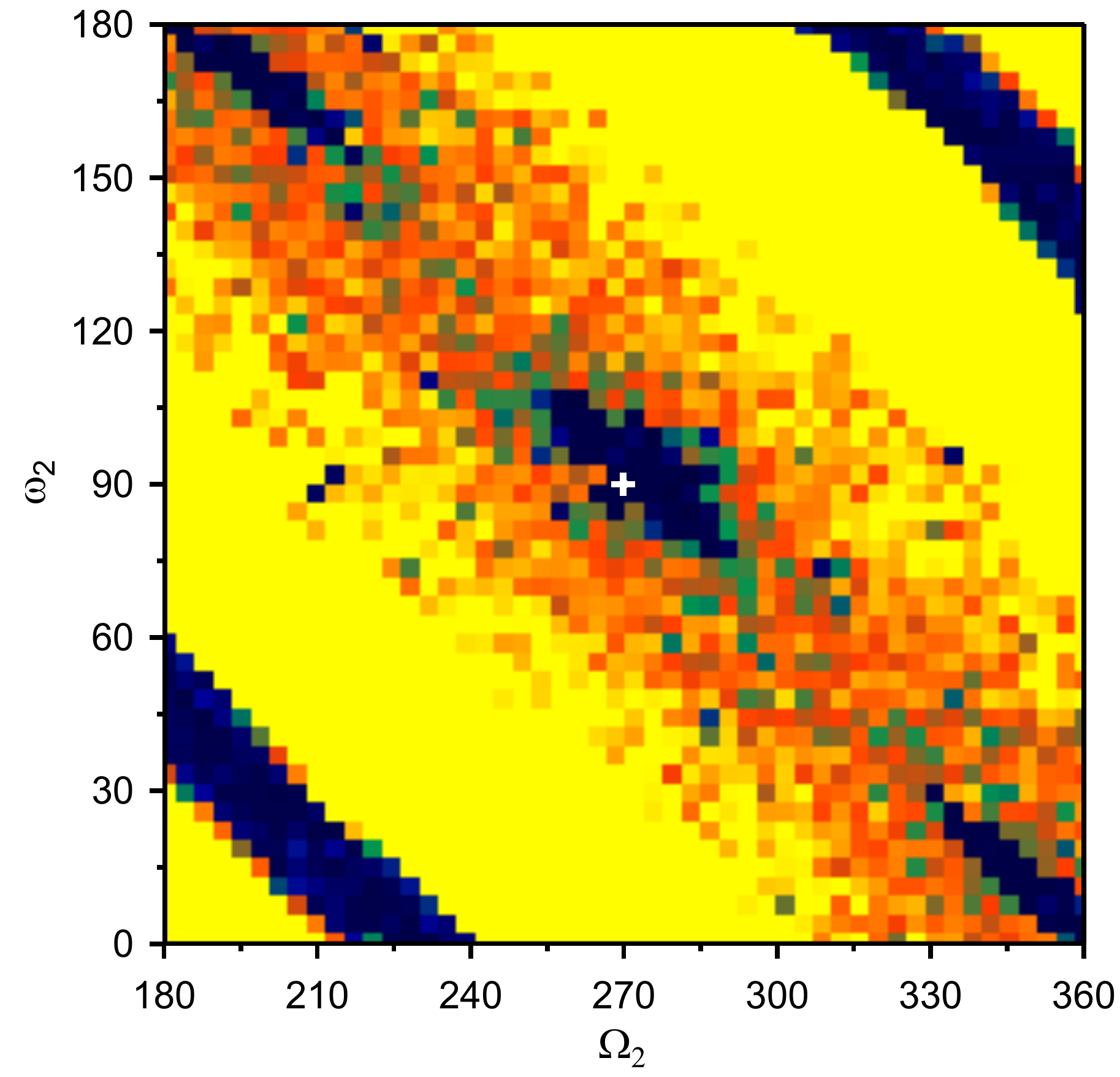}  \\
\textnormal{(c)}  & \;&\textnormal{(d)} \\
\end{array} $
$\begin{array}{c}
\includegraphics[width=3.5cm,height=.6cm]{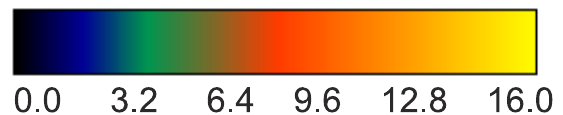}
\end{array} $
\end{center}
\caption{DS-maps on the {\bf a} $(e_1,i_2)$, {\bf b} $(\Omega_2,i_2)$, {\bf c} $(\Omega_2,M_2)$ and {\bf d} $(\Omega_2,\varpi_2)$ plane.}
\label{maps}
\end{figure}

In Fig. \ref{maps}, we present DS-maps in the neighbourhood of the stable spatial 1/1 resonant periodic orbit shown in Fig. \ref{inrot} (see also \cite{hpv09,hv1111,robu,avv14} and references therein). The grids are of size $50\times 50$ and of different combinations of the orbital elements of the planets. We compute the \textit{de-trended} Fast Lyapunov Indicator (see \cite{froe97,voyatzis08}) for a maximum time of $200$Ky. Dark coloured regions represent the orbits with long-term stability and the white crosses represent the stable periodic orbit.

\begin{figure}
\begin{center}
$\begin{array}{ccc}
\includegraphics[width=0.45\textwidth]{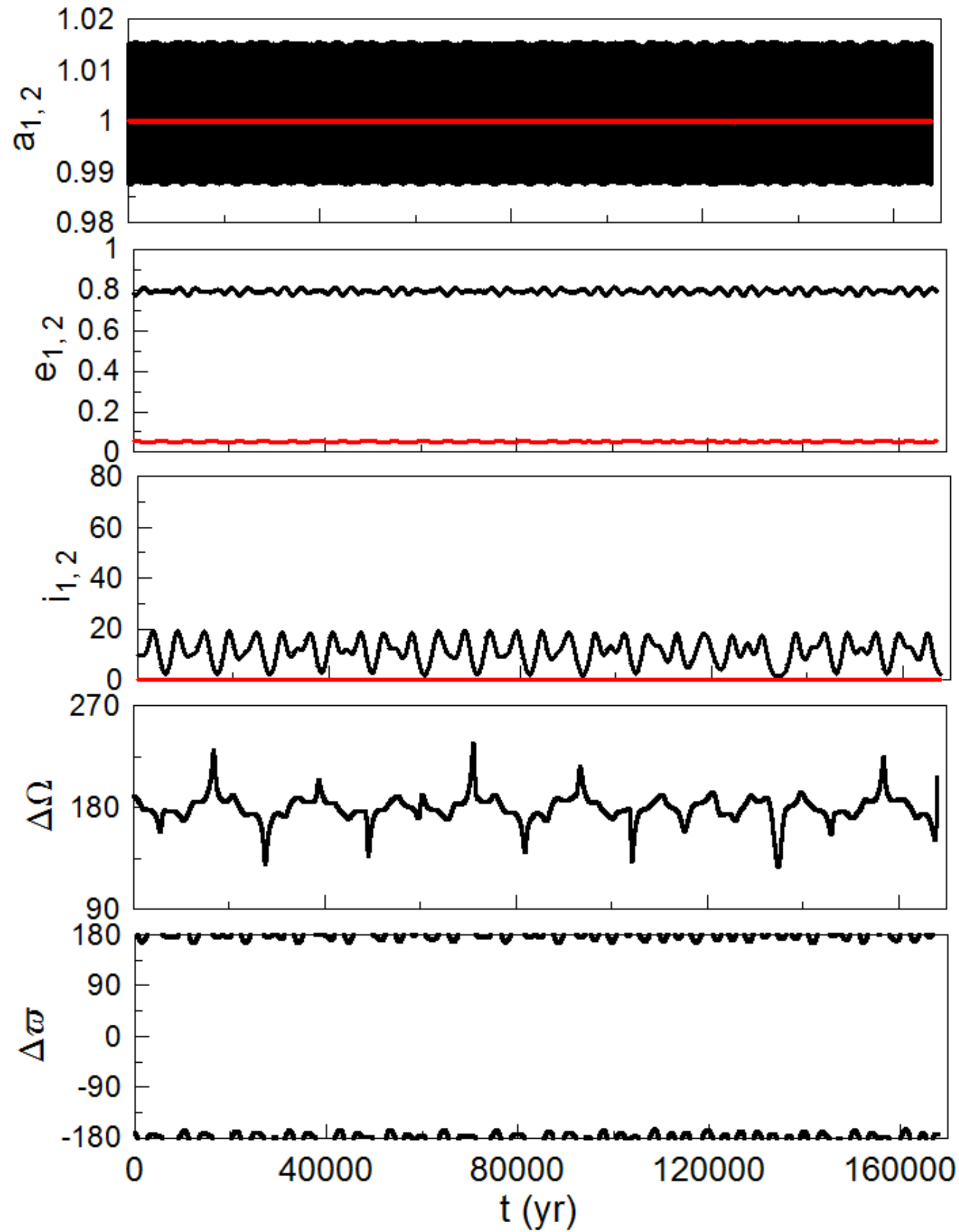}  &\;&
\includegraphics[width=0.45\textwidth]{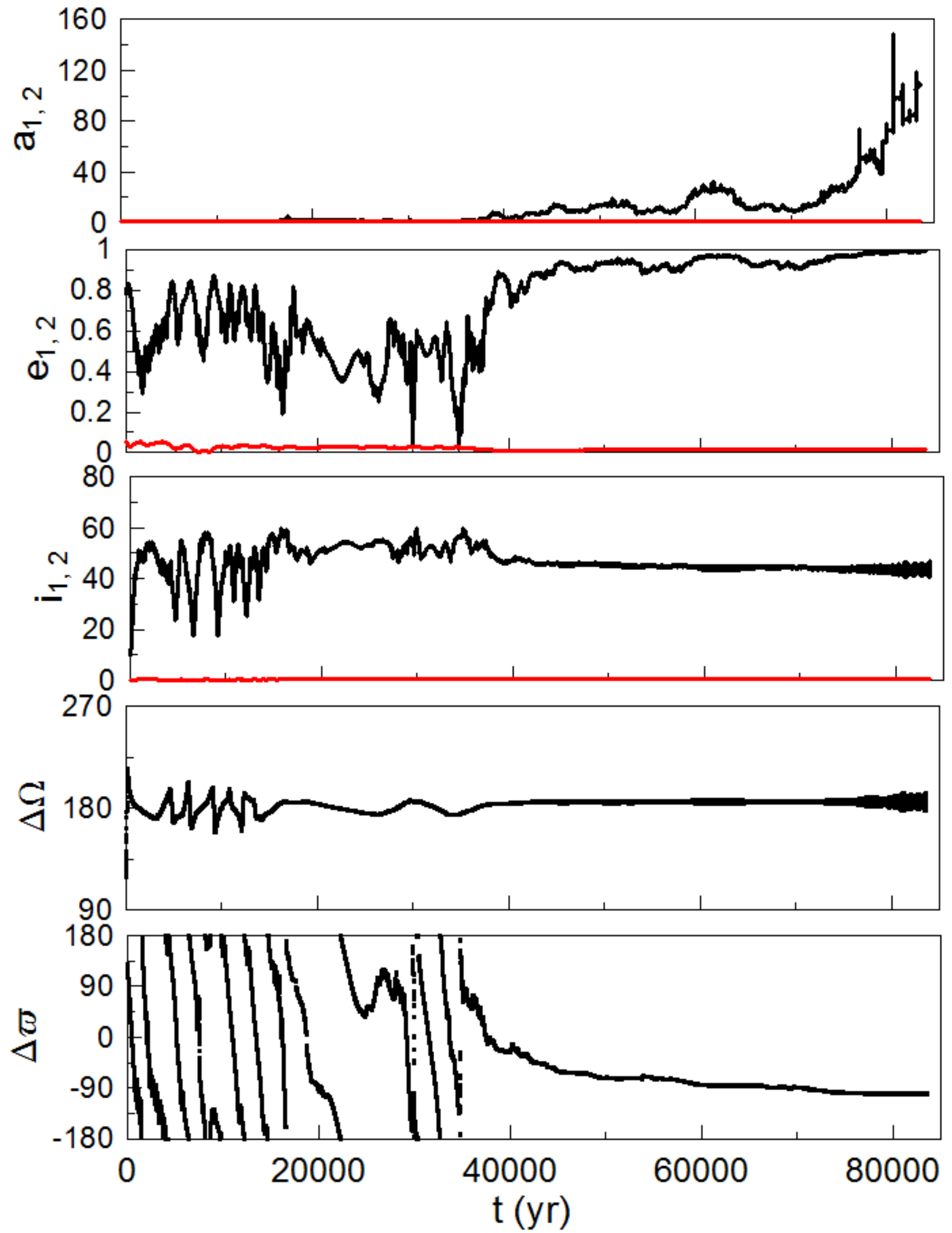}  \\
\textnormal{(a)}  & \;&\textnormal{(b)} 
\end{array} $
\end{center}
\caption{Evolution in the neighbourhood of the stable periodic orbit of Fig. \ref{inrot}. Based on Fig. \ref{maps}c and altering $M_2$ and $\Omega_2$ we observe {\bf a} regular ($M_2=15^\circ, \Omega_2=280^\circ$) and {\bf b} chaotic orbits ($M_2=-33^\circ, \Omega_2=220^\circ$). The outer planet is scattered in the latter case.}
\label{11evol}
\end{figure}

In Fig. \ref{11evol}, we exemplify the location of regular and chaotic orbits in phase space given a stable periodic orbit.

\section{Evolution of resonant exoplanets}\label{sec8}

We show the relation between the long-term evolution of extrasolar planetary systems trapped in MMRs and the evolution in the vicinity of resonant stable periodic orbits discussed previously. The equations of motion in the rotating frame were solved numerically with a minimum accuracy $10^{-14}$ by using the Bulirsch-Stoer integrator. The periodic orbits were computed with an accuracy of 12 decimal digits after successive differential approximation of the periodicity conditions. Throughout the study, we normalized the total mass of the system and the semi-major axis of the inner planet, $a_1$, to unity ($a_2$ was computed, in order to satisfy the resonance studied in each case). 

\subsection{HD 82943}\label{sec9}

We study the system HD $82943$(b,c) of masses $m_1=0.004018224$, $m_2=0.004035037$ and $m_0=0.991946$, whose planets are locked in 2/1 MMR and use the initial conditions \cite{tan13}: 
\begin{equation}
\begin{array}{llll}
a_1=1.00,&e_1=0.425,&\varpi_1=133^{\circ},&M_1=256^{\circ}\\
a_2=1.5951,&e_2=0.203,&\varpi_2=107^{\circ},&M_2=333^{\circ}
\label{hd82943ic}
\end{array}
\end{equation}

In Fig. \ref{hd82943f}, we observe regular libration around the initial values (\ref{hd82943ic}) of the orbital elements and the angles  $0^{\circ}$, $0^{\circ}$ and $0^{\circ}$ of the resonant angles $\theta_1$, $\Delta\varpi$ and $\theta_2$, respectively. Thus, this system can be considered as stable evolving around the stable symmetric periodic orbit of the family of mass ratio $\rho=1.004$ in the configuration $(\theta_1, \theta_2)=(0,0)$ shown in Fig. \ref{pos}a. The periodic orbit is linearly both horizontally and vertically stable (the v.c.o. is depicted by the magenta coloured dot) and consequently, these exoplanets could survive, if they were mutually inclined. Moreover, such an evolution in the neighbourhood of a planar stable periodic orbit can only exhibit e-resonance, namely the resonant angles indicating i-resonance will not librate. These resonant angles librate, iff the evolution takes place in the neighbourhood of a spatial periodic orbit (see e.g. \cite{vat14}). In Fig. \ref{mps}a, we depict the island of stability which hosts this planetary system.

\subsection{HD 73526}\label{sec10}

We examine the system HD $73526$(b,c) of masses $m_1 =0.0021668$, $m_2 =0.0020192$ and $m_0=0.995813$, whose planets are evolving in 2/1 MMR and use the initial conditions \cite{wit14}: 
\begin{equation}
\begin{array}{llll}
a_1=1.00,&e_1=0.244,&\varpi_1=198.3^{\circ},&M_1=105^{\circ}\\
a_2=1.5846,&e_2=0.169,&\varpi_2=294.5^{\circ},&M_2=153.4^{\circ}
\label{hd73526ic}
\end{array}
\end{equation}

\begin{figure*}[h]
\centering
\begin{tabular}{cc}
\begin{minipage}[t]{.45\textwidth}
  \includegraphics[width=\textwidth,keepaspectratio]{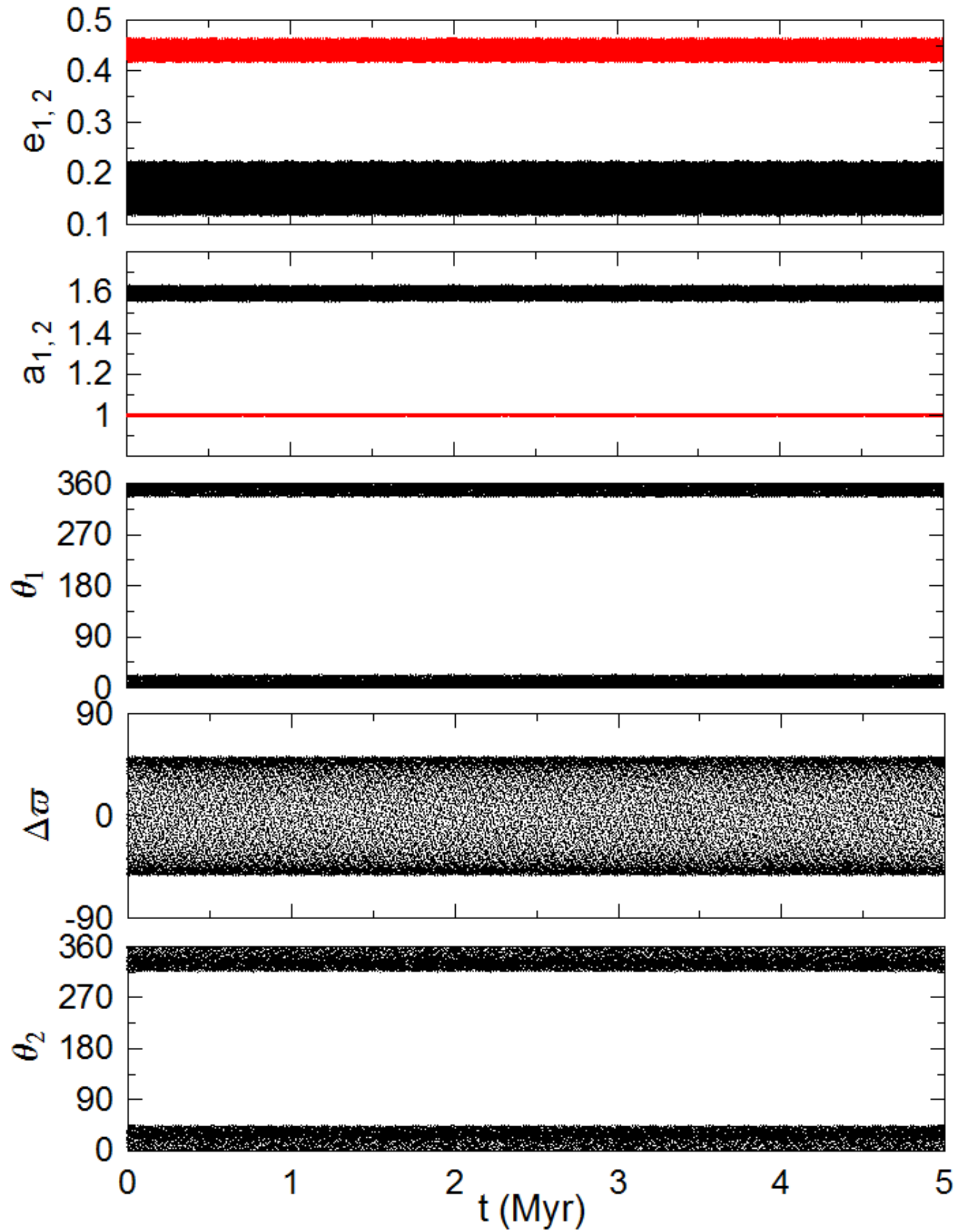}
  \caption{The evolution of orbital elements $a_i$, $e_i$, resonant angles $\theta_1$, $\Delta\varpi$ and $\theta_2$ of the planetary system HD $82943$ with initial conditions \eqref{hd82943ic}. Red and black line stands for inner ($P_1$) and outer planet, $P_2$, respectively.}
  \label{hd82943f}
\end{minipage}%
& 
\begin{minipage}[t]{.45\textwidth}
  \includegraphics[width=\textwidth,keepaspectratio]{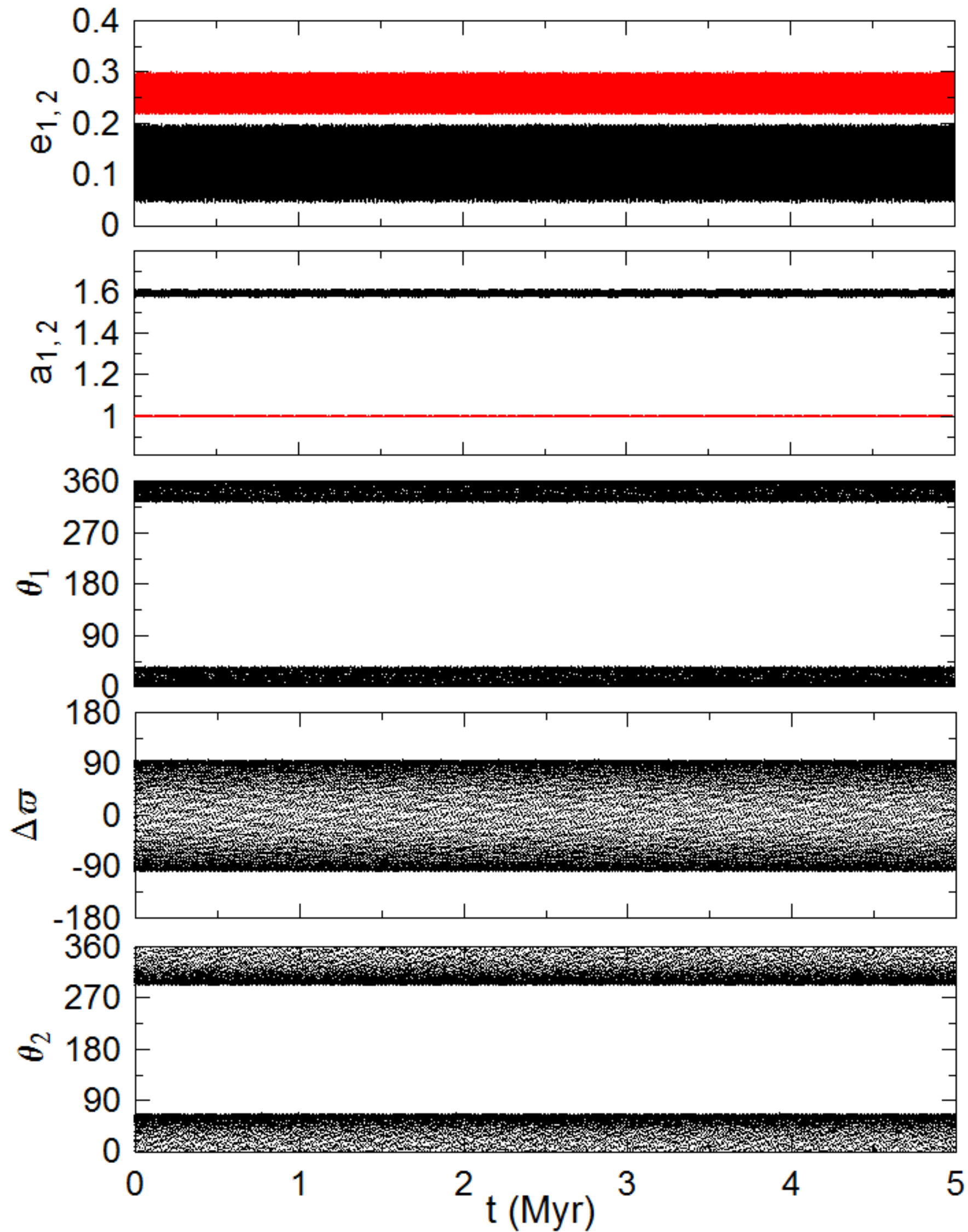}
  \caption{The evolution of the planetary system HD $73526$ with initial conditions (\ref{hd73526ic}) (presentation as in Fig. \ref{hd82943f}).}
  \label{hd73526f}
\end{minipage}\end{tabular}
\end{figure*}

Due to the regular librations shown in Fig. \ref{hd73526f}, this system can be considered evolving around a planar asymmetric periodic orbit of the family of mass ratio $\rho=0.86$ generated by a symmetric periodic orbit in the configuration $(\theta_1, \theta_2)=(0,0)$ (see also Fig. \ref{pos}b). The computed DS-map around this asymmetric periodic orbit (where the evolution is precessing about) revealed the hosting island of stability (see Fig. \ref{mps}b). Regarding the resonant angles, there exists an island of stability, around this asymmetric periodic orbit, which is centered at $\Delta M=180^{\circ}$ and $\Delta \varpi =0$ and hence, confirms the observed oscillations. Likewise HD $82943$, this periodic orbit is linearly both horizontally and vertically stable and in the context described in Sect. \ref{sec9} the planets could be mutually inclined. In \cite{long}, simulations of arbitrarily chosen initial conditions in the neighbourhood of the observational data have been performed and they found -although not disrupted for 10Gy- irregular evolutions. It is the stable periodic orbit that should guide the extent of deviations from the orbital elements, in order for regular evolutions to be found.

\subsection{HD 128311}\label{sec11}

We consider the system HD $128311$(b,c) of masses $m_1 =0.0025786$, $m_2 =0.003797$ and $m_0 = 0.993624$, whose planets are trapped to 2/1 MMR and use the initial conditions \cite{vogt05}: 
\begin{equation}
\begin{array}{llll}
a_1=1.00,&e_1=0.25,&\varpi_1=110.9^{\circ},&M_1=159.4^{\circ}\\
a_2=1.601455,&e_2=0.17,&\varpi_2=195.5^{\circ},&M_2=0^{\circ}
\label{hd128311ic}
\end{array}
\end{equation}

In Fig. \ref{hd128311f}, we observe that the orbital elements evolve regularly around their initial values (\ref{hd128311ic}), while the  resonant angles, $\theta_1$, $\Delta\varpi$ and $\theta_2$ librate around the angles $0^{\circ}$, $0^{\circ}$ and $0^{\circ}$, respectively. Thus, this system can be considered as stable evolving around the symmetric family of mass ratio $\rho=1.47$ in the configuration $(\theta_1, \theta_2)=(0,0)$ (Figs. \ref{pos}c,\ref{mps}c). Similarly to the previously described systems, the evolution takes place about a linearly both horizontally and vertically stable periodic orbit. Thus, initial conditions for a regular evolution of a non-planar configuration (only irregular evolutions were reported in \cite{long} for arbitrarily chosen initial conditions) could only exist in the vicinity of the planar stable symmetric periodic orbit. 
 
\subsection{HD 60532}\label{sec12}

\begin{figure*}[h]
\centering
\begin{tabular}{cc}
\begin{minipage}[t]{.45\textwidth}
  \includegraphics[width=\textwidth,keepaspectratio]{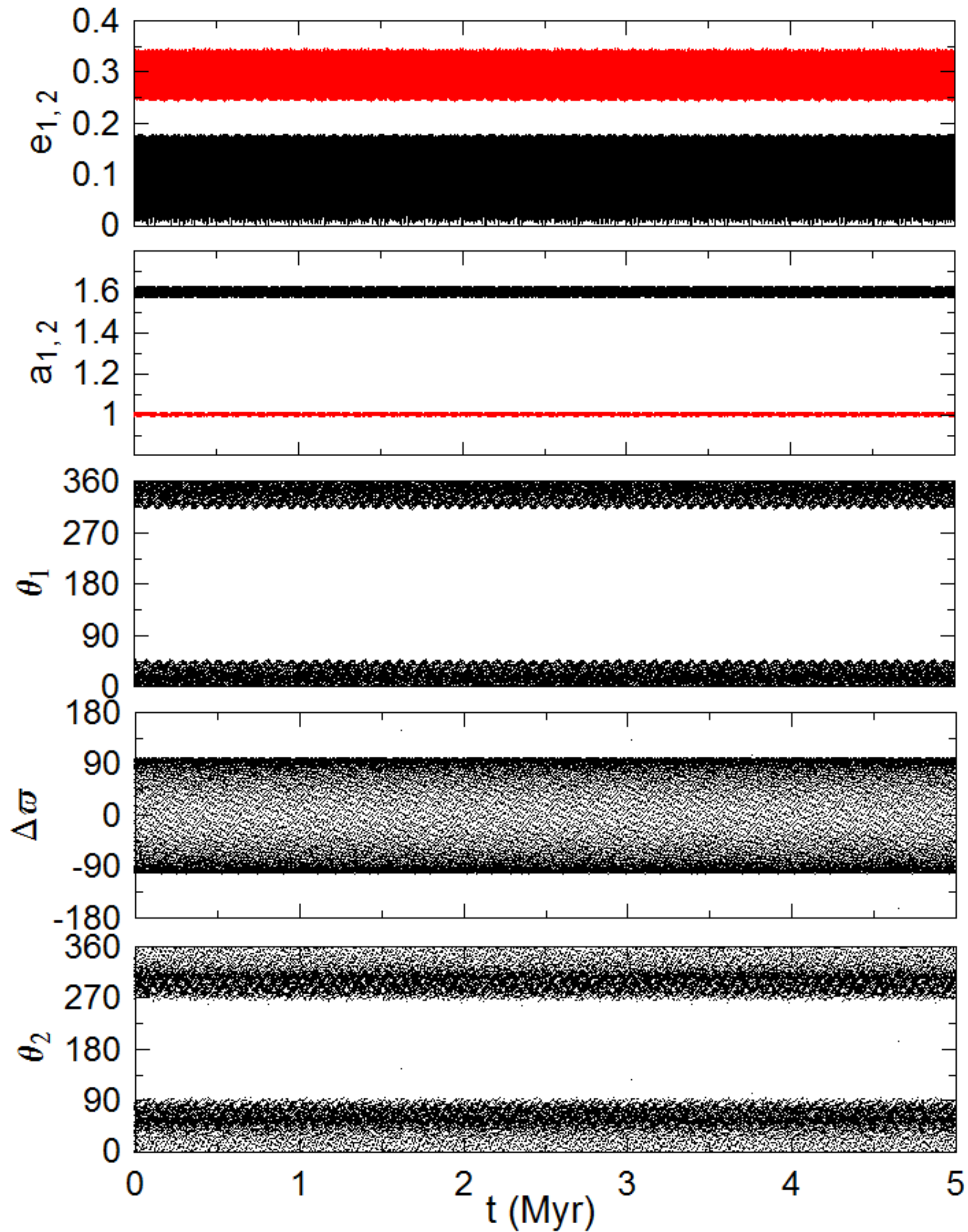}
  \caption{The evolution of HD $128311$ with initial conditions (\ref{hd128311ic}) (presented as in Fig. \ref{hd82943f}).}
  \label{hd128311f}
\end{minipage}%
& 
\begin{minipage}[t]{.45\textwidth}
  \includegraphics[width=\textwidth,keepaspectratio]{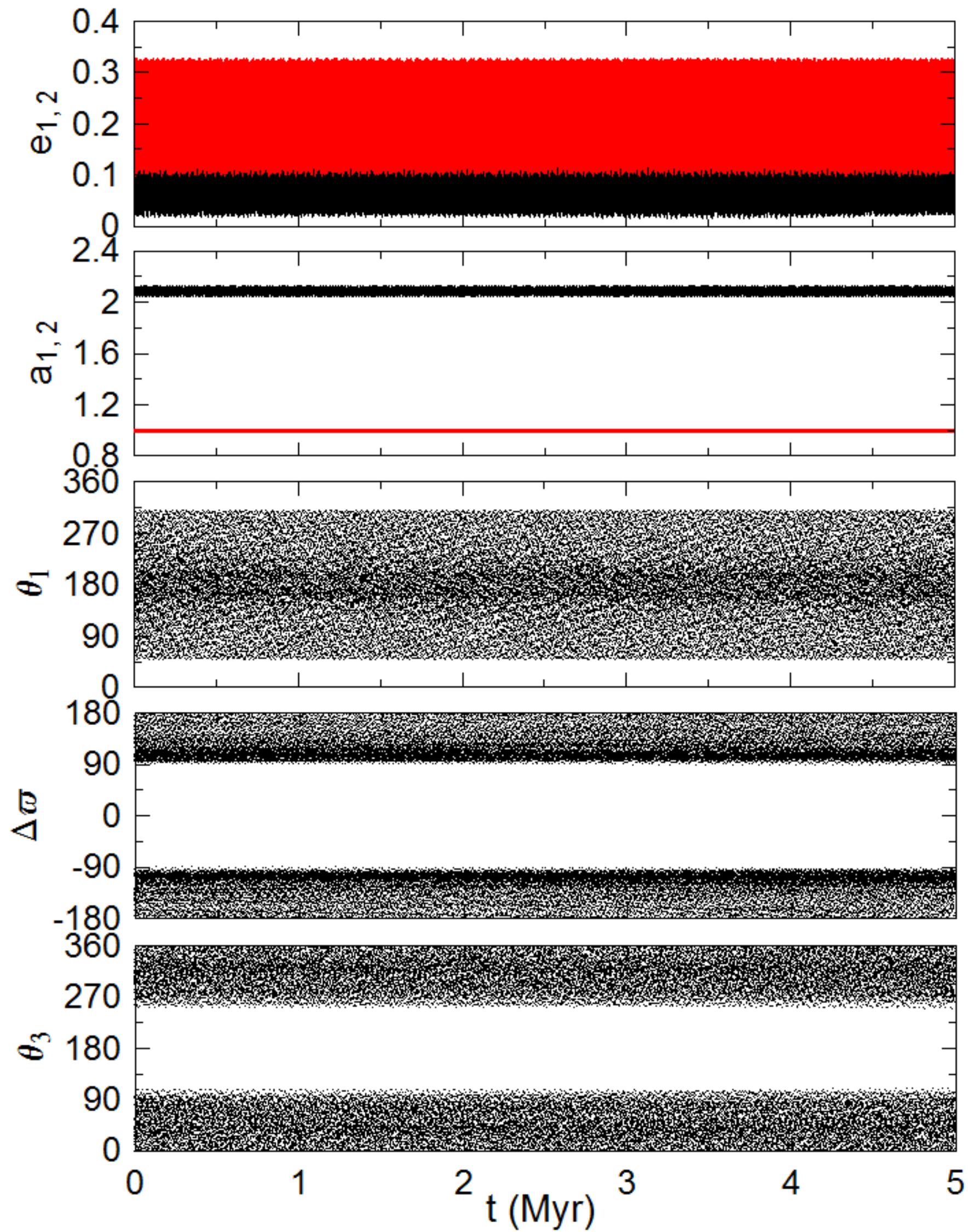}
  \caption{The evolution of  HD $60532$ with initial conditions (\ref{hd60532ic}) (presented as in Fig. \ref{hd82943f}).}
  \label{hd60532f}
\end{minipage}
\end{tabular}
\end{figure*}

We examine the system HD $60532$(b,c) of masses $m_1=0.0021748$, $m_2=0.0051450$ and $m_0=0.99268$, whose planets are captured to 3/1 MMR and use the initial conditions \cite{lask09}: 
\begin{equation}
\begin{array}{llll}
a_1=1.00,&e_1=0.278,&\varpi_1=352.83^{\circ},&M_1=21.95^{\circ}\\
a_2=2.0844,&e_2=0.038,&\varpi_2=119.49^{\circ},&M_2=197.53^{\circ}
\label{hd60532ic}
\end{array}
\end{equation}
In Fig. \ref{hd60532f}, we observe that the orbital elements evolve regularly around their initial values (\ref{hd60532ic}), while the resonant angles, $\theta_1$, $\Delta\varpi$ and $\theta_3$ librate around the angles $180^{\circ}$, $180^{\circ}$ and $0^{\circ}$, respectively. Thus, this system can be considered as stable evolving around the symmetric family of mass ratio $\rho=2.36$ in the configuration $(\theta_3, \theta_1)=(0,\pi)$ (Figs. \ref{pos}d,\ref{mps}d). The stability of this system was, also, verified by \cite{lask09}. Regarding non-planar regular configurations, sought in \cite{long}, the linearly horizontally and vertically stable symmetric periodic orbit can guide the initial conditions chosen in its vicinity.

\subsection{HD 45364}\label{sec13}

We analyze the system HD $45364$(b,c) of masses $m_1=0.0002280$, $m_2=0.0008014$ and $m_0=0.99897$, whose planets are locked in 3/2 MMR and use the initial conditions \cite{cor09}: 
\begin{equation}
\begin{array}{llll}
a_1=1.00,&e_1=0.1684,&\varpi_1=162.58^{\circ},&M_1=-56.82^{\circ}\\
a_2=1.3168,&e_2=0.0974,&\varpi_2=7.41^{\circ},&M_2=262.11^{\circ}
\label{hd45364ic}
\end{array}
\end{equation}

In Fig. \ref{hd45364f}, we observe that the orbital elements evolve regularly around their initial values (\ref{hd45364ic}), while the resonant angles, $\theta_1$, $\Delta\varpi$ and $\theta_2$ librate around the angles $0^{\circ}$, $180^{\circ}$ and $180^{\circ}$, respectively (also confirmed by \cite{cor09}). Thus, this system can be considered as stable evolving exactly centered at a stable symmetric periodic orbit of the family of mass ratio $\rho=3.51$ in the configuration $(\theta_1, \theta_2)=(0,0)$ (Figs. \ref{pos}e,\ref{mps}e). The stable non-planar configuration found in \cite{long} is due to the existence of a horizontally and vertically stable periodic orbit in the vicinity of the initial conditions chosen.

\subsection{HD 108874}\label{sec14}

We consider the system HD $108874$(b,c) of masses $m_1=0.00136$, $m_2 = 0.0010183$ and $m_0=0.997622$ and use the initial conditions \cite{vogt05}: 
\begin{equation}
\begin{array}{llll}
a_1=1.00,&e_1=0.07,&\varpi_1=248.4^{\circ},&M_1=83.13^{\circ}\\
a_2=2.549952,&e_2=0.32,&\varpi_2=17.3^{\circ},&M_2=0^{\circ}
\label{hd108874ic}
\end{array}
\end{equation}

\begin{figure*}[h]
\centering
\begin{tabular}{cc}
\begin{minipage}[t]{.45\textwidth}
  \includegraphics[width=\textwidth,keepaspectratio]{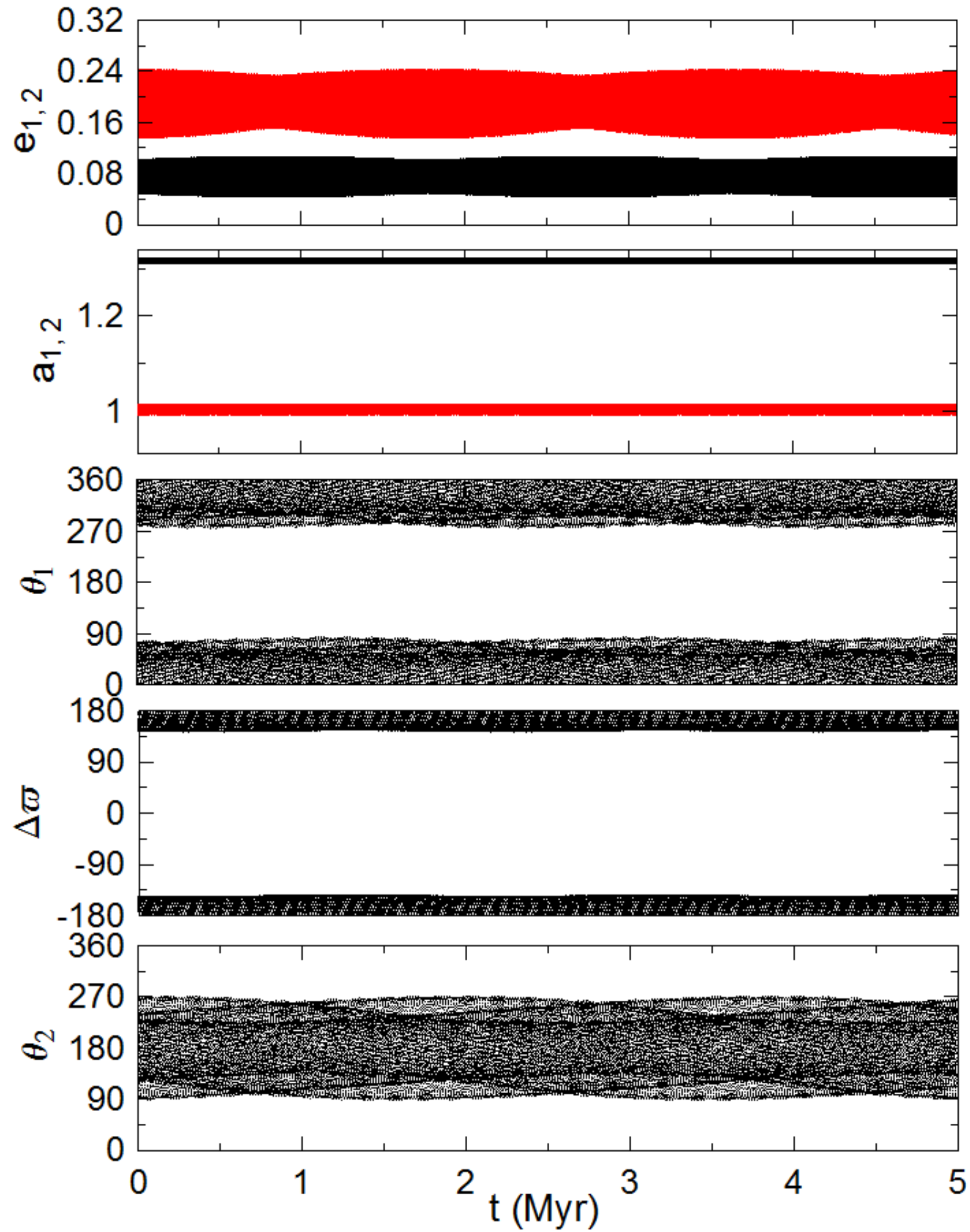}
  \caption{The evolution of  HD $45364$ with initial conditions (\ref{hd45364ic}) (presented as in Fig. \ref{hd82943f}).}
  \label{hd45364f}
\end{minipage}%
& 
\begin{minipage}[t]{.45\textwidth}
  \includegraphics[width=\textwidth,keepaspectratio]{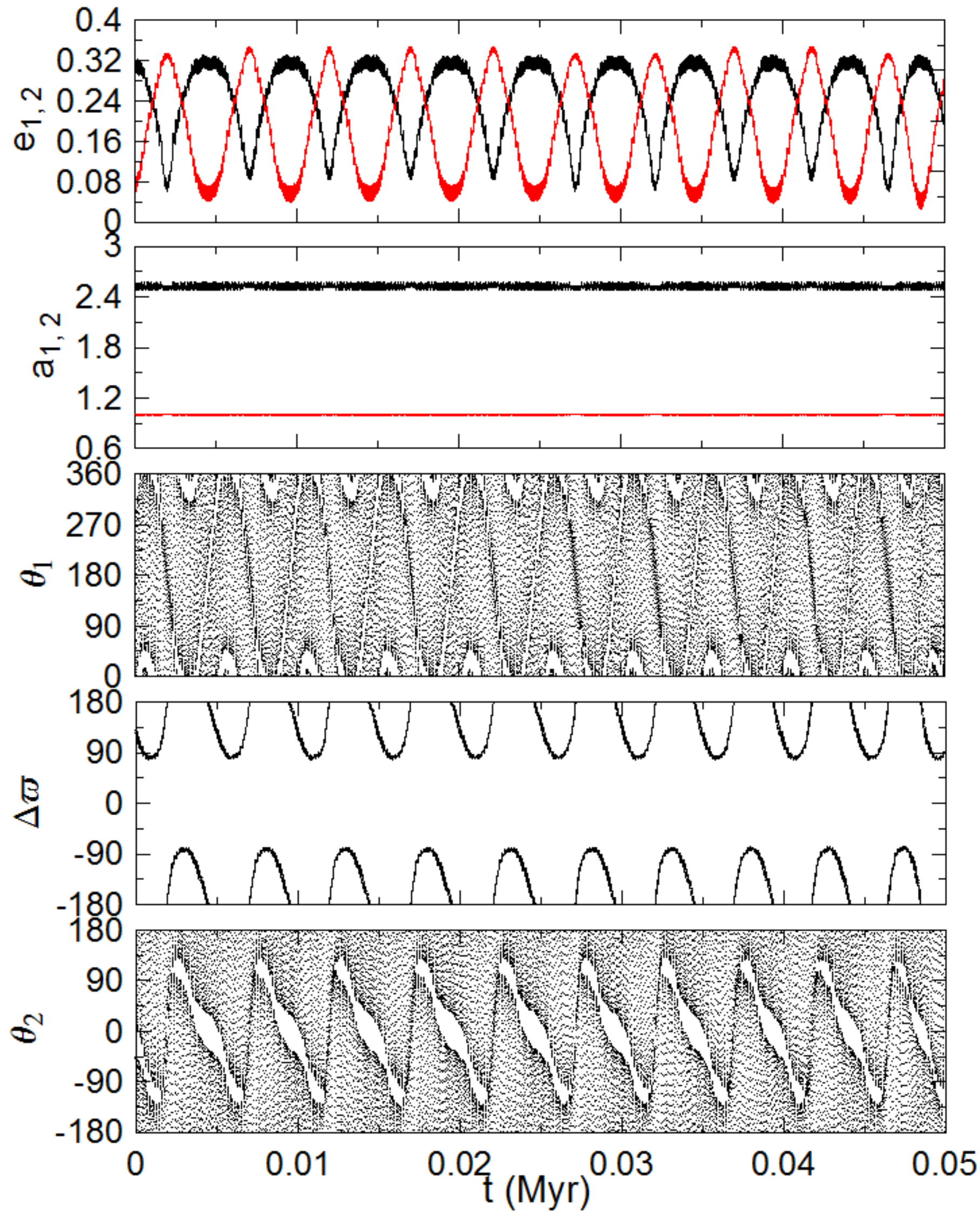}
  \caption{The evolution of HD $108874$ with initial conditions (\ref{hd108874ic}) shown for a small amount of time (presented as in Fig. \ref{hd82943f}).}
  \label{hd108874f}
\end{minipage}
\end{tabular}
\end{figure*}

\begin{figure*}[h]
\begin{center}
$\begin{array}{ccc}
\includegraphics[width=5.5cm,height=5.5cm]{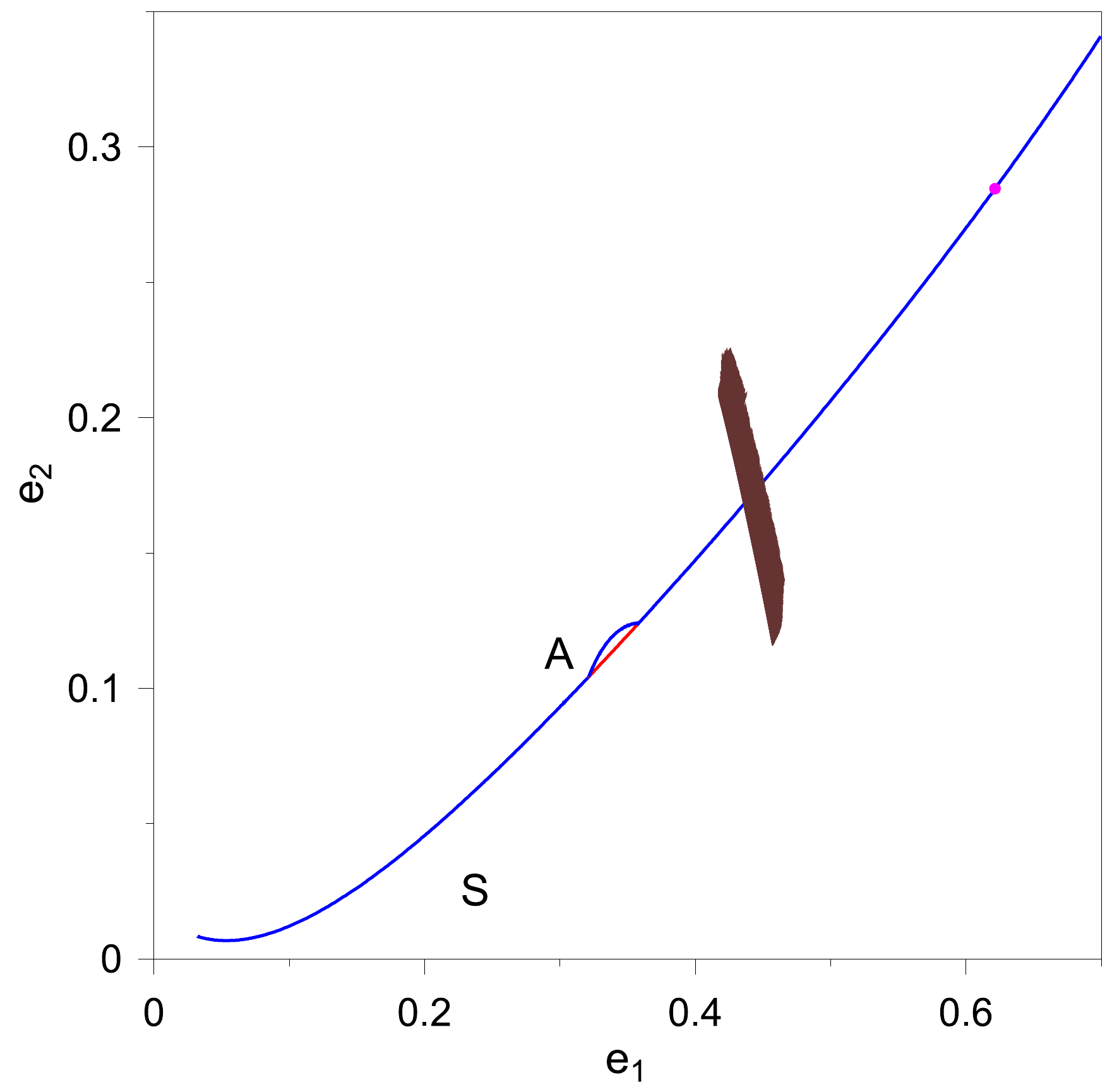}  &\;&\includegraphics[width=5.5cm,height=5.5cm]{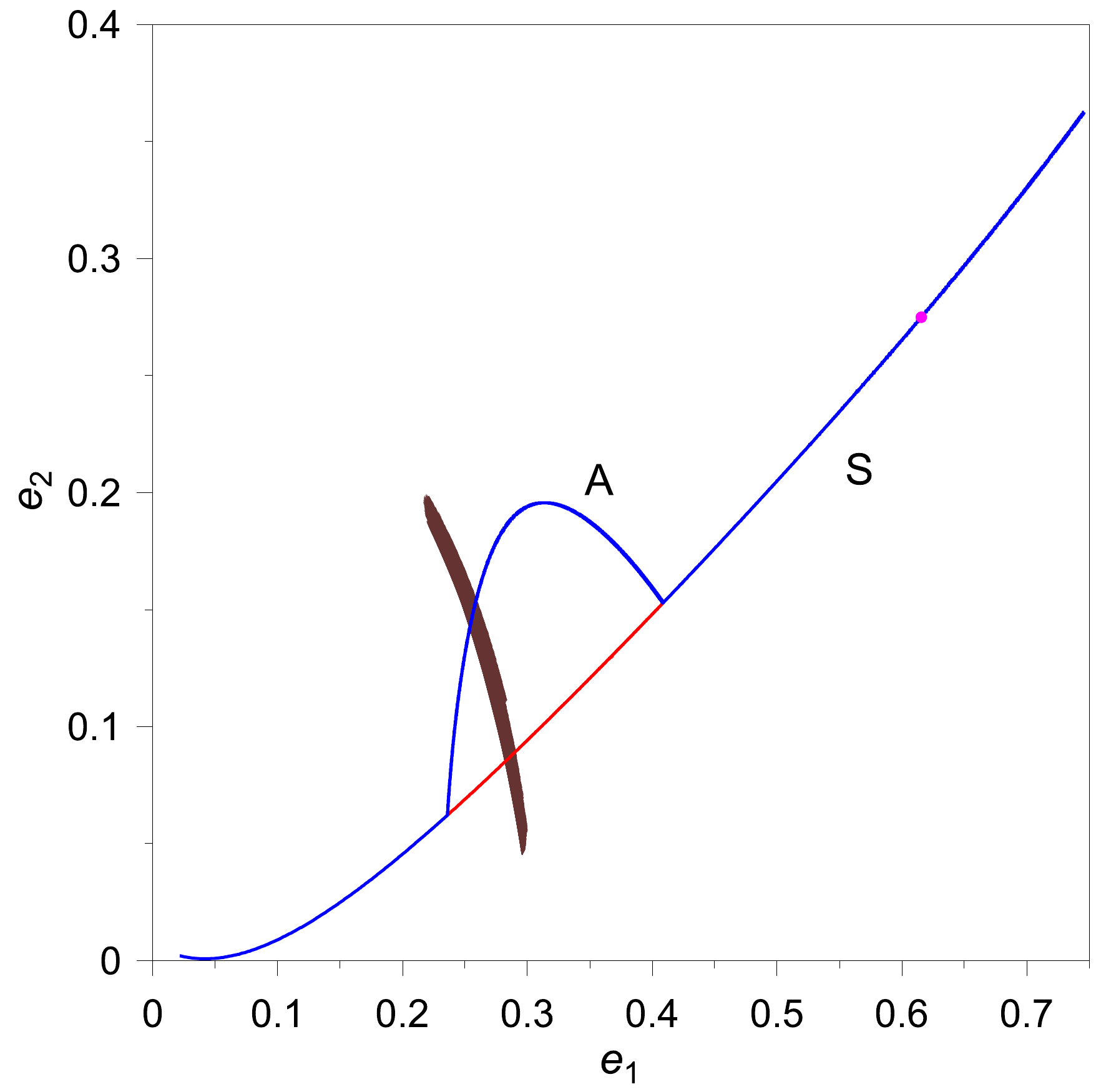}  \\
\textnormal{(a)}  & \;&\textnormal{(b)} \\
\includegraphics[width=5.5cm,height=5.5cm]{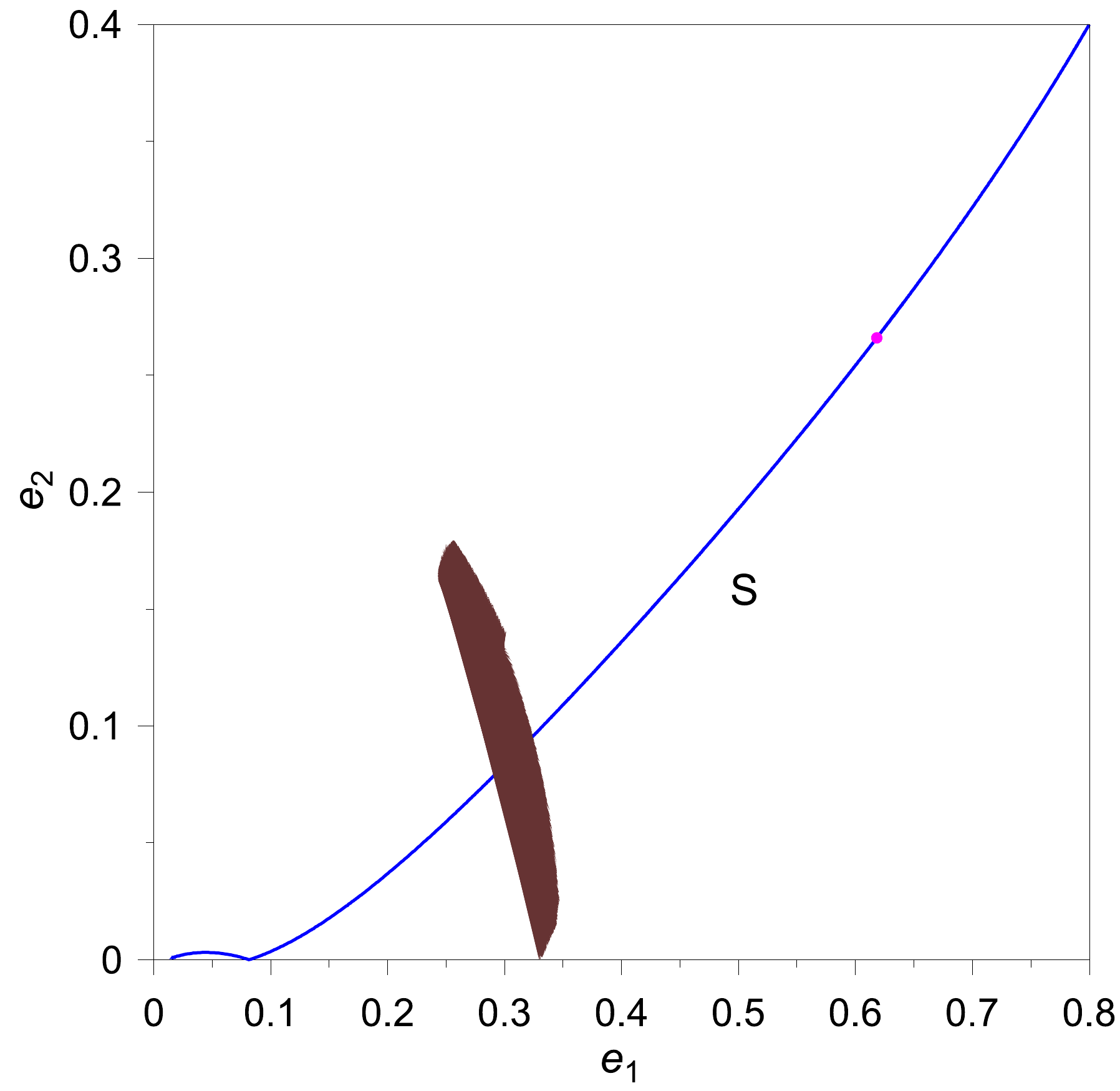} &\;& \includegraphics[width=5.5cm,height=5.5cm]{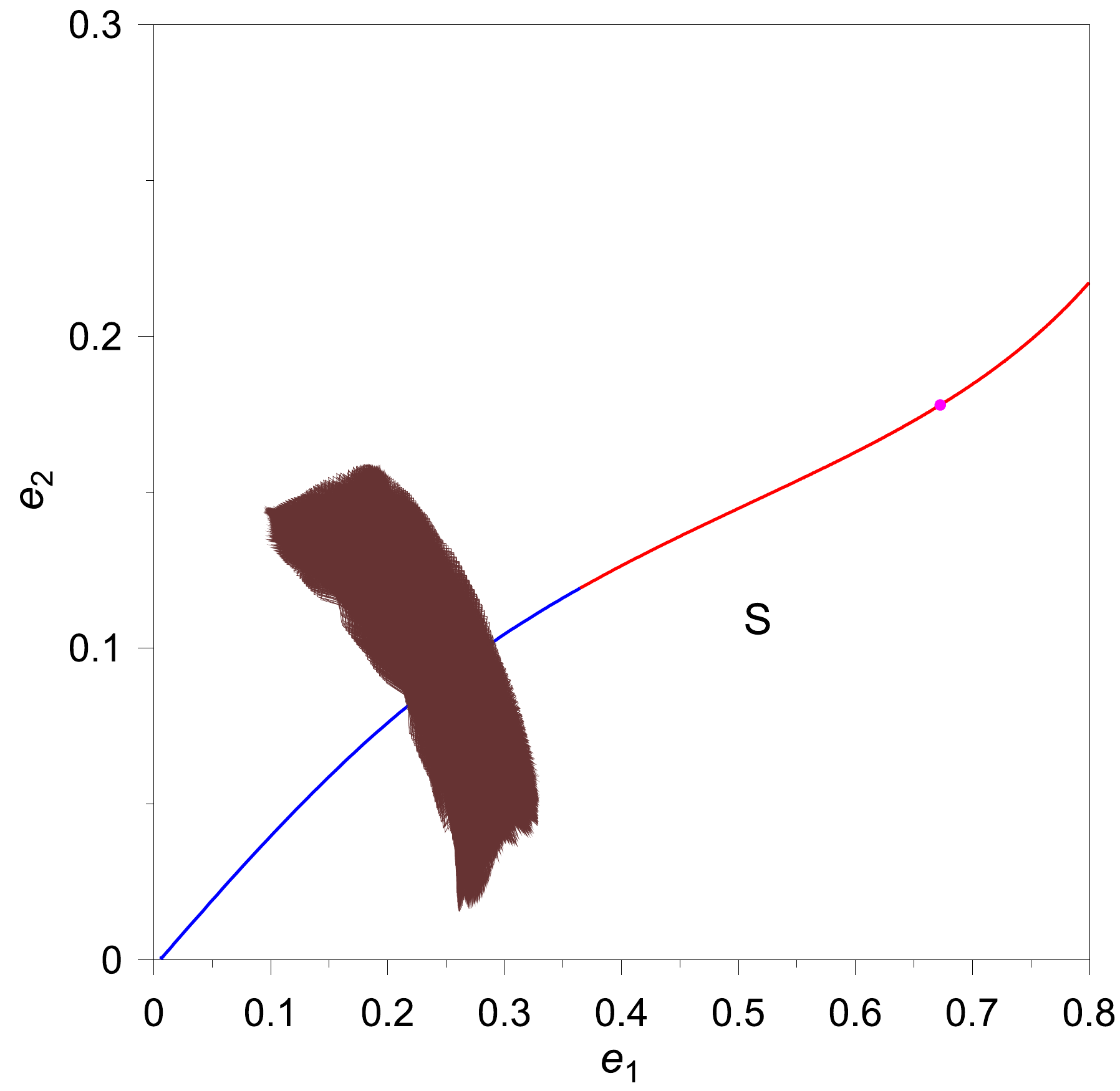}  \\
\textnormal{(c)}  & \;&\textnormal{(d)} \\
\includegraphics[width=5.5cm,height=5.5cm]{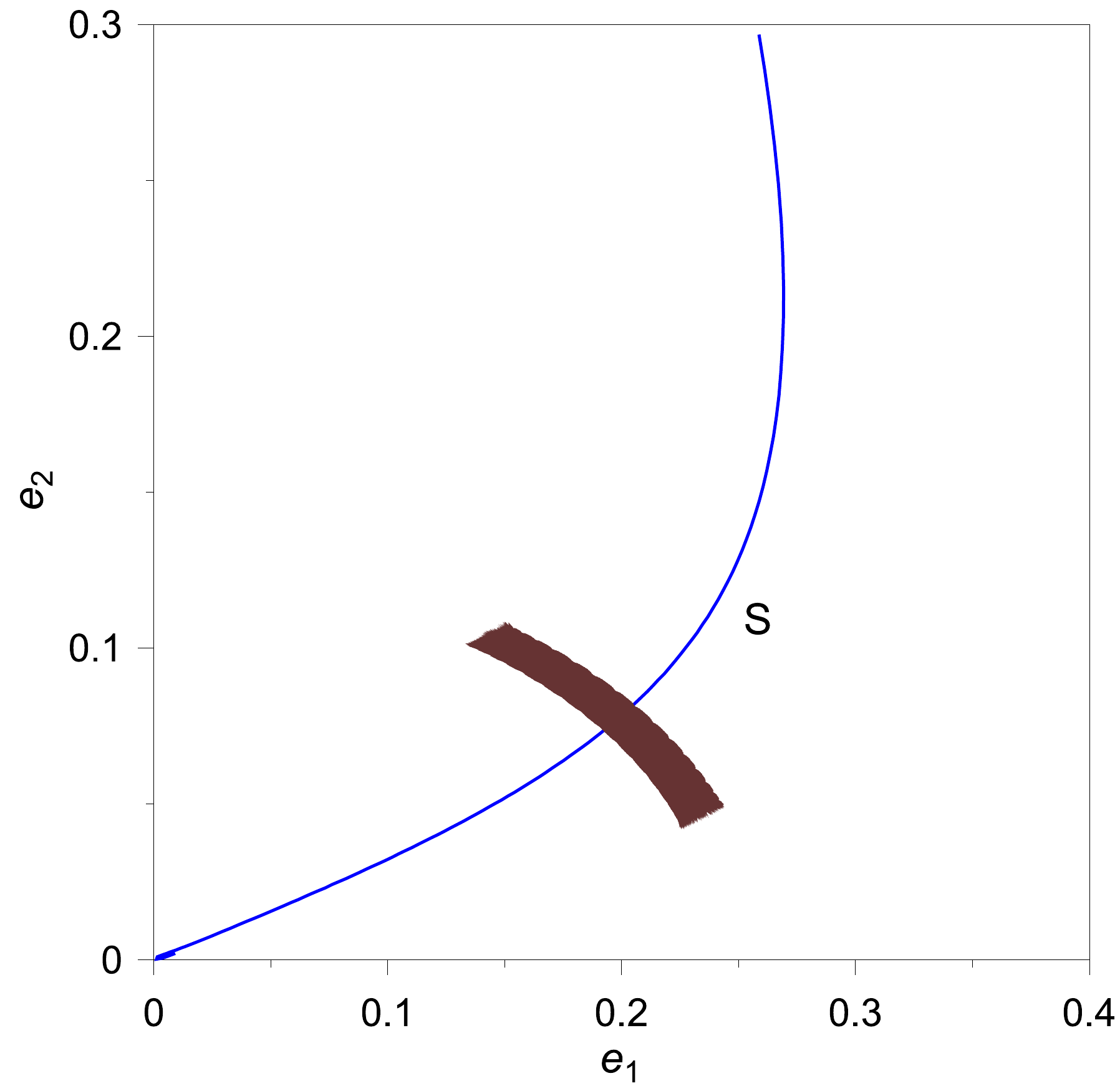} &\;& \includegraphics[width=5.5cm,height=5.5cm]{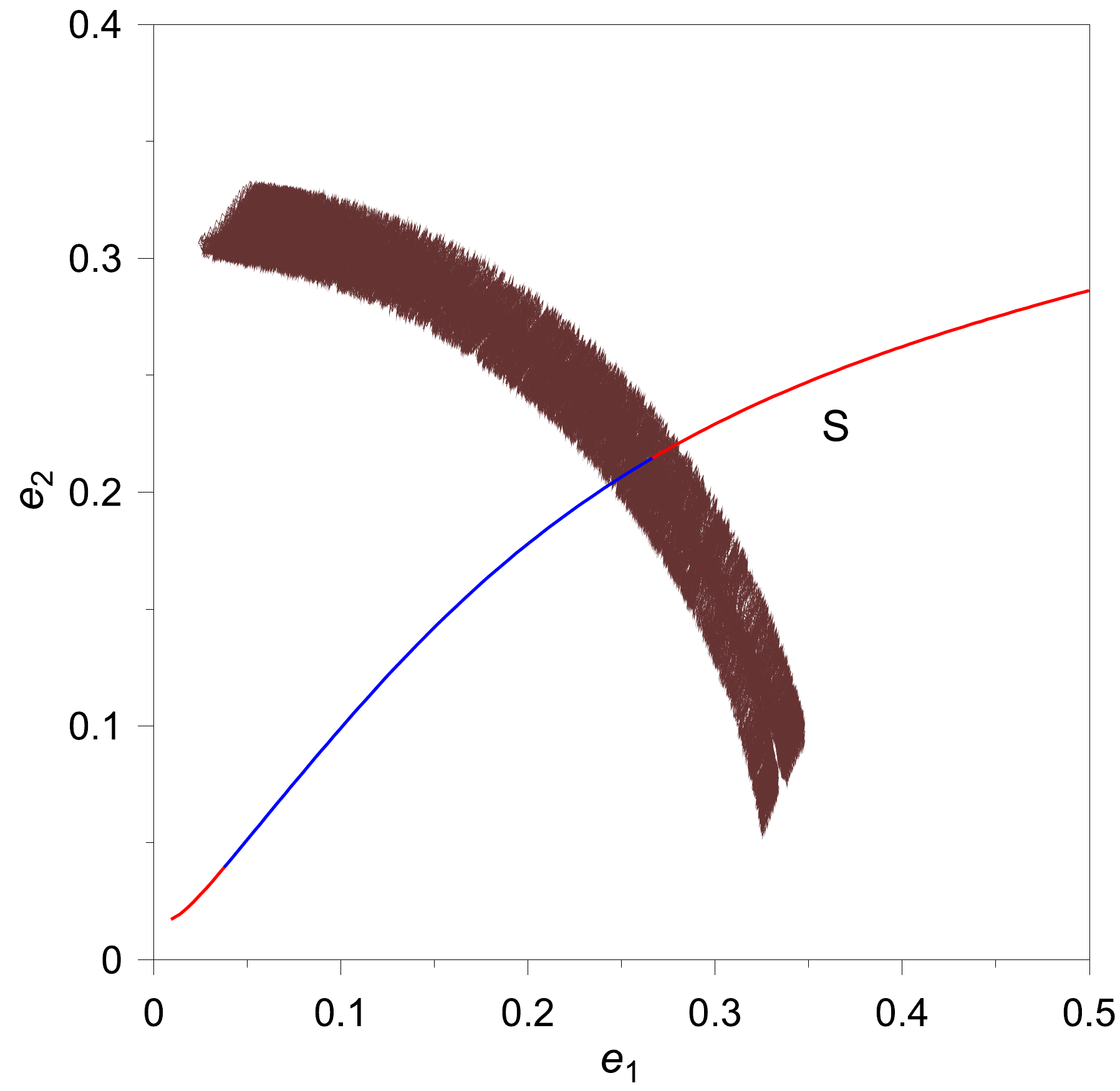}  \\
\textnormal{(e)}  & \;&\textnormal{(f)} \\
\end{array} $\vspace{-2em}
\end{center}
\caption{The evolution (brown) of {\bf a} HD $82943$, {\bf b} HD $73526$, {\bf c} HD $128311$,  {\bf d} HD $60532$, {\bf e} HD $45364$ and {\bf f} HD $108874$ shown for 5 Myr,  exactly centered at a stable periodic orbit. The v.c.o. is depicted by a magenta coloured dot. Up to that orbit the planar orbits are vertically stable. Blue and red lines correspond to horizontally stable and unstable orbits. A and S depict the asymmetric and symmetric periodic orbits, respectively.}
\label{pos}
\end{figure*}

\begin{figure*}[h]
\begin{center}
$\begin{array}{ccc}
\includegraphics[width=5.5cm,height=5.5cm]{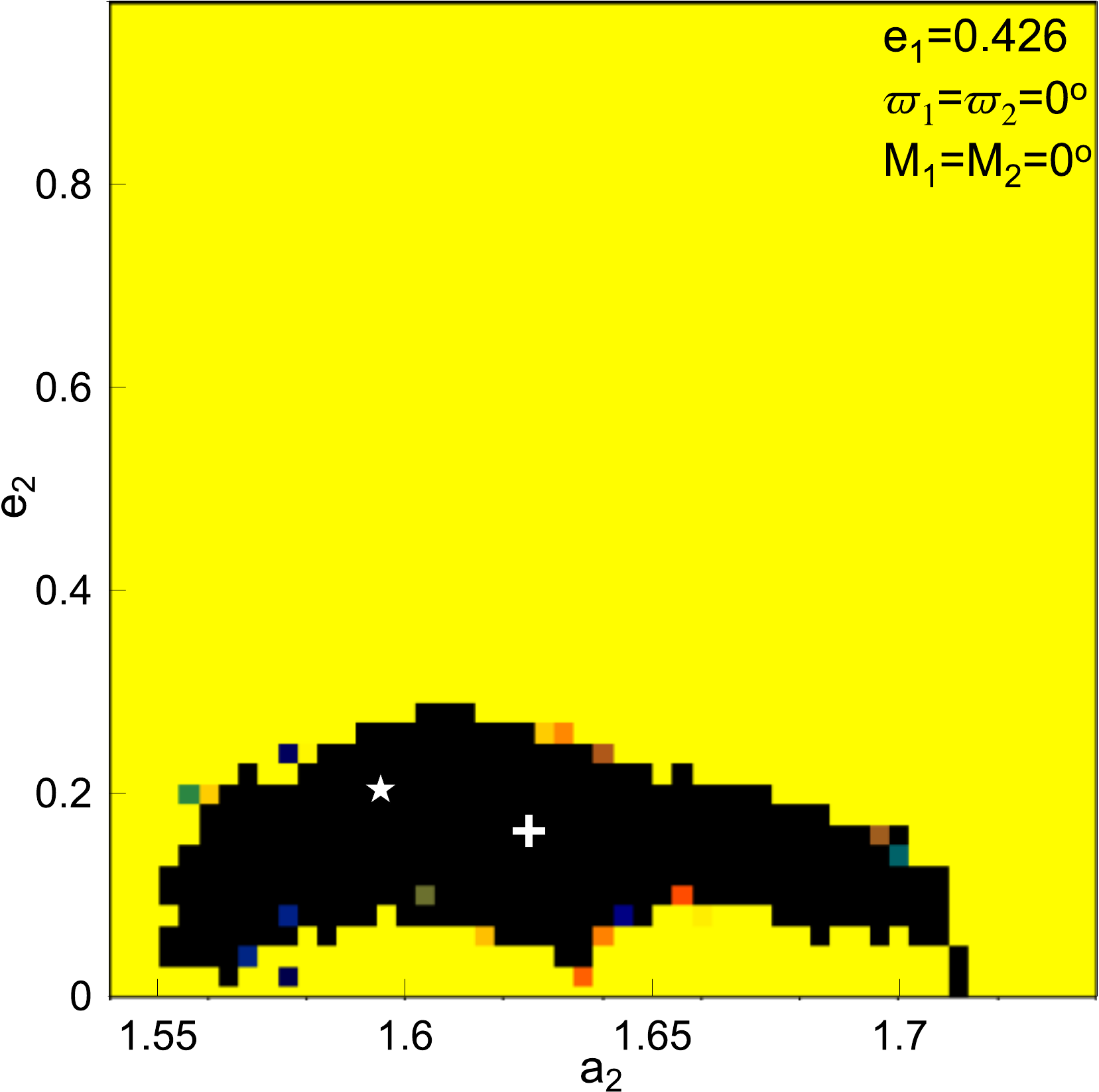}  &\;&\includegraphics[width=5.5cm,height=5.5cm]{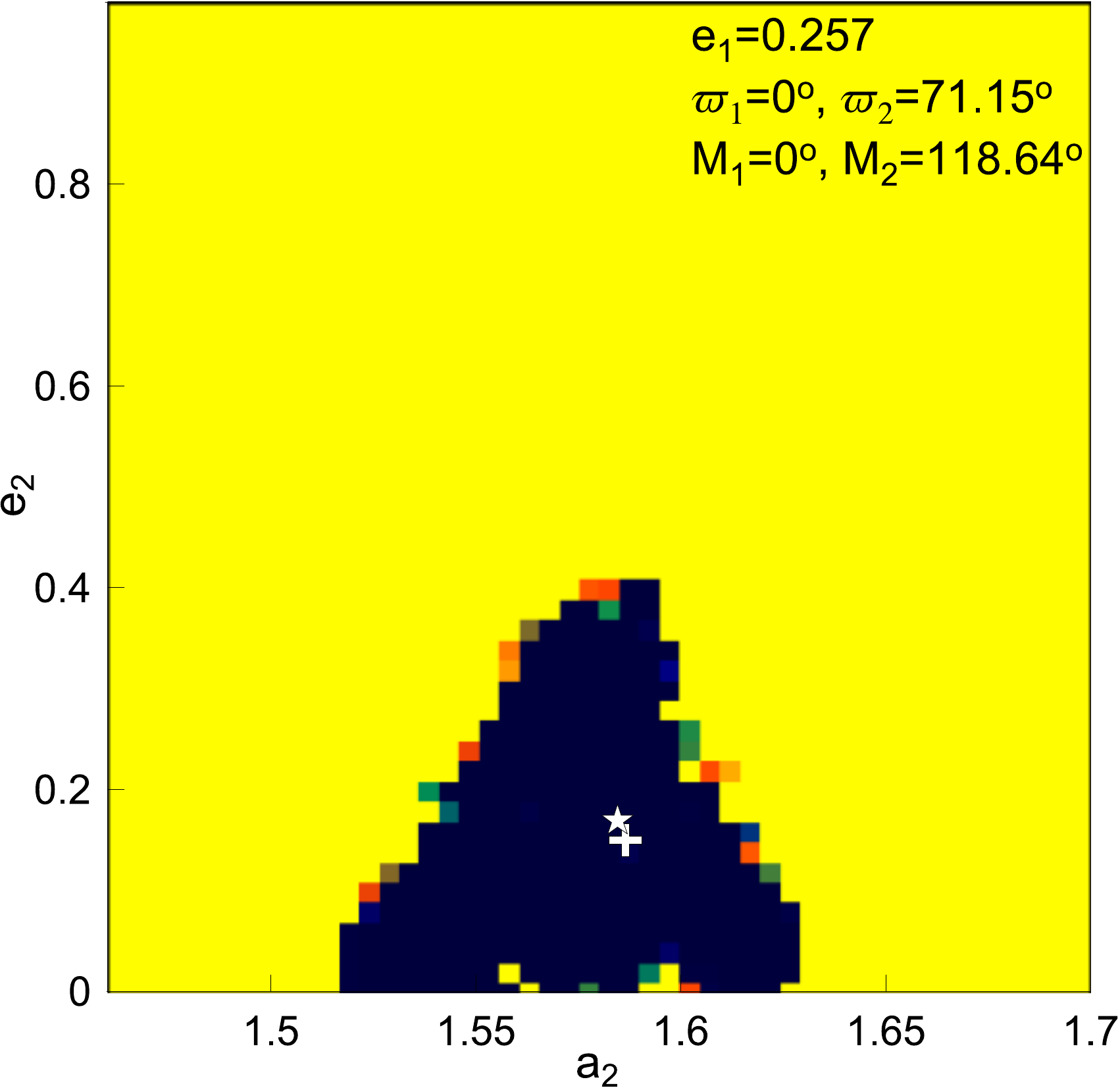}  \\
\textnormal{(a)}  & \;&\textnormal{(b)} \\
\includegraphics[width=5.5cm,height=5.5cm]{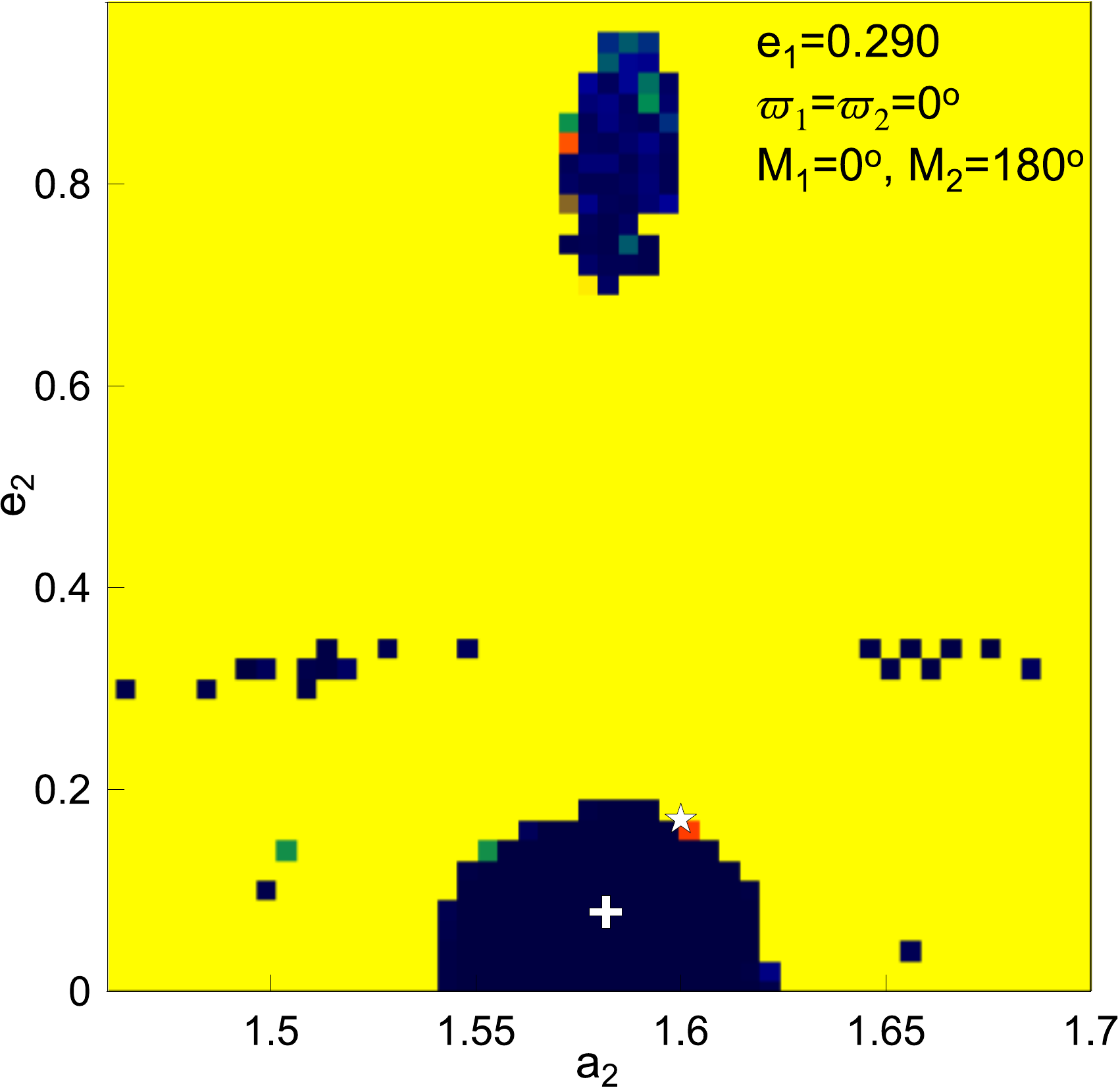} &\;& \includegraphics[width=5.5cm,height=5.5cm]{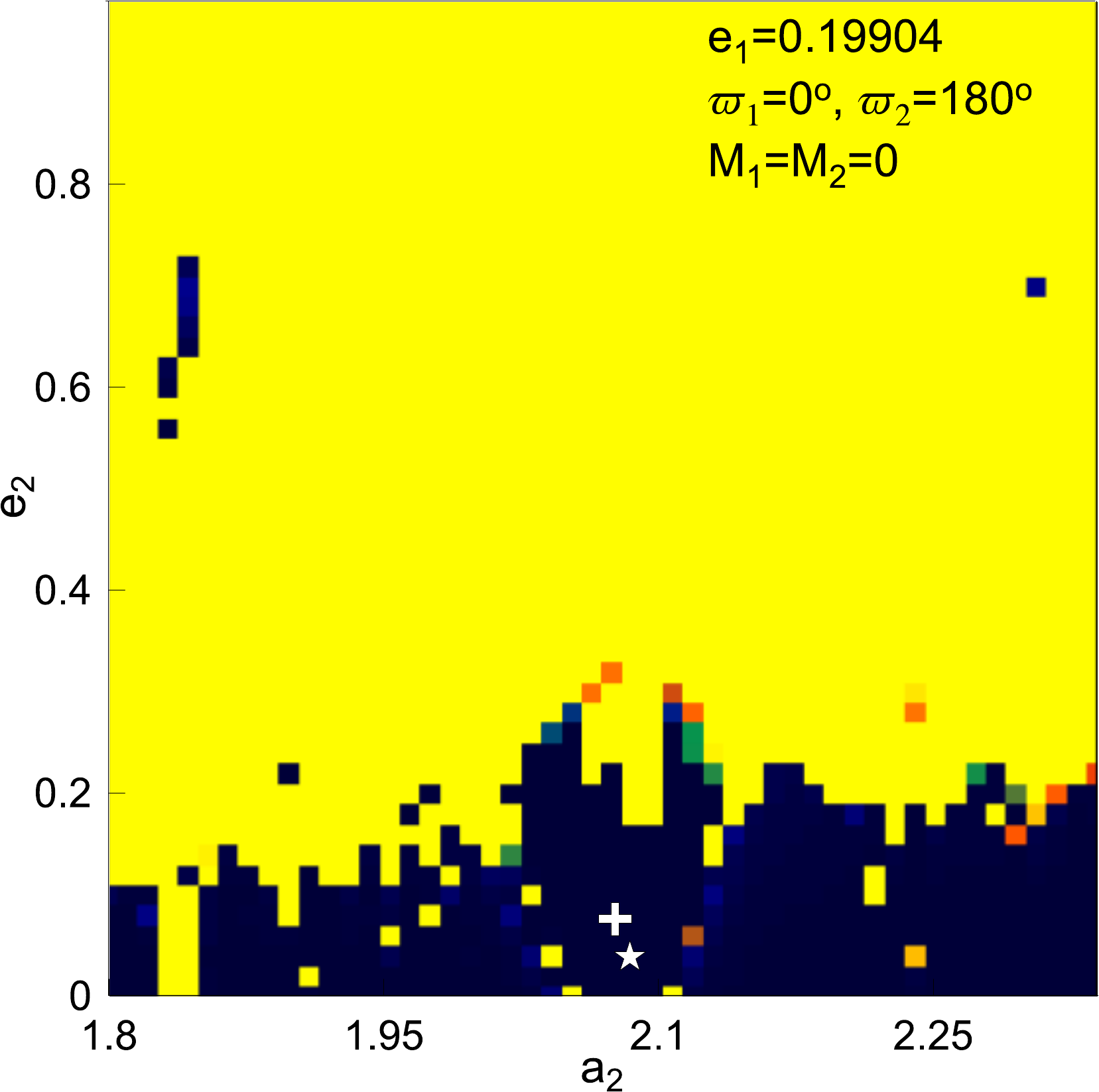}  \\
\textnormal{(c)}  & \;&\textnormal{(d)} \\
\includegraphics[width=5.5cm,height=5.5cm]{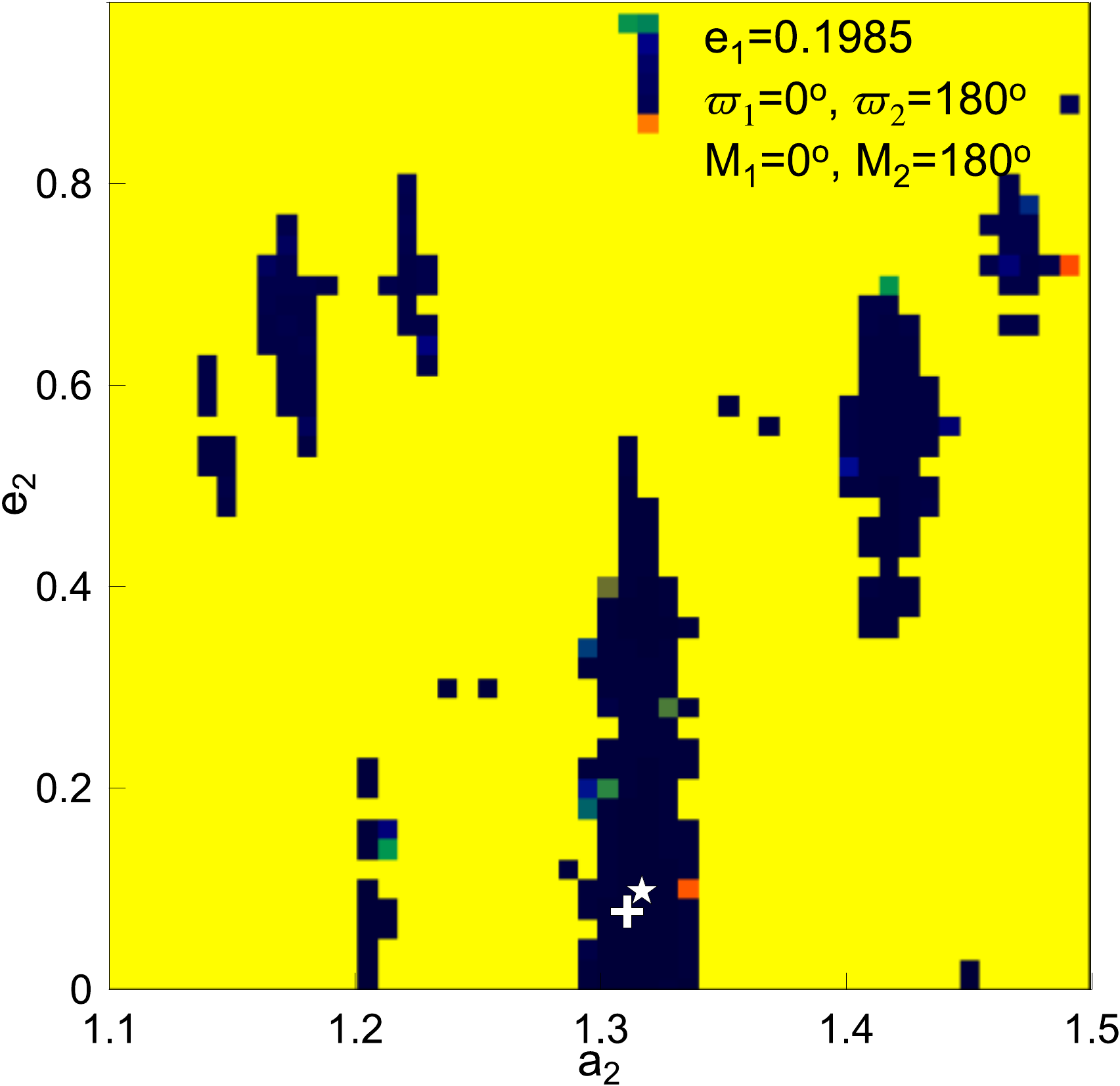} &\;& \includegraphics[width=5.5cm,height=5.5cm]{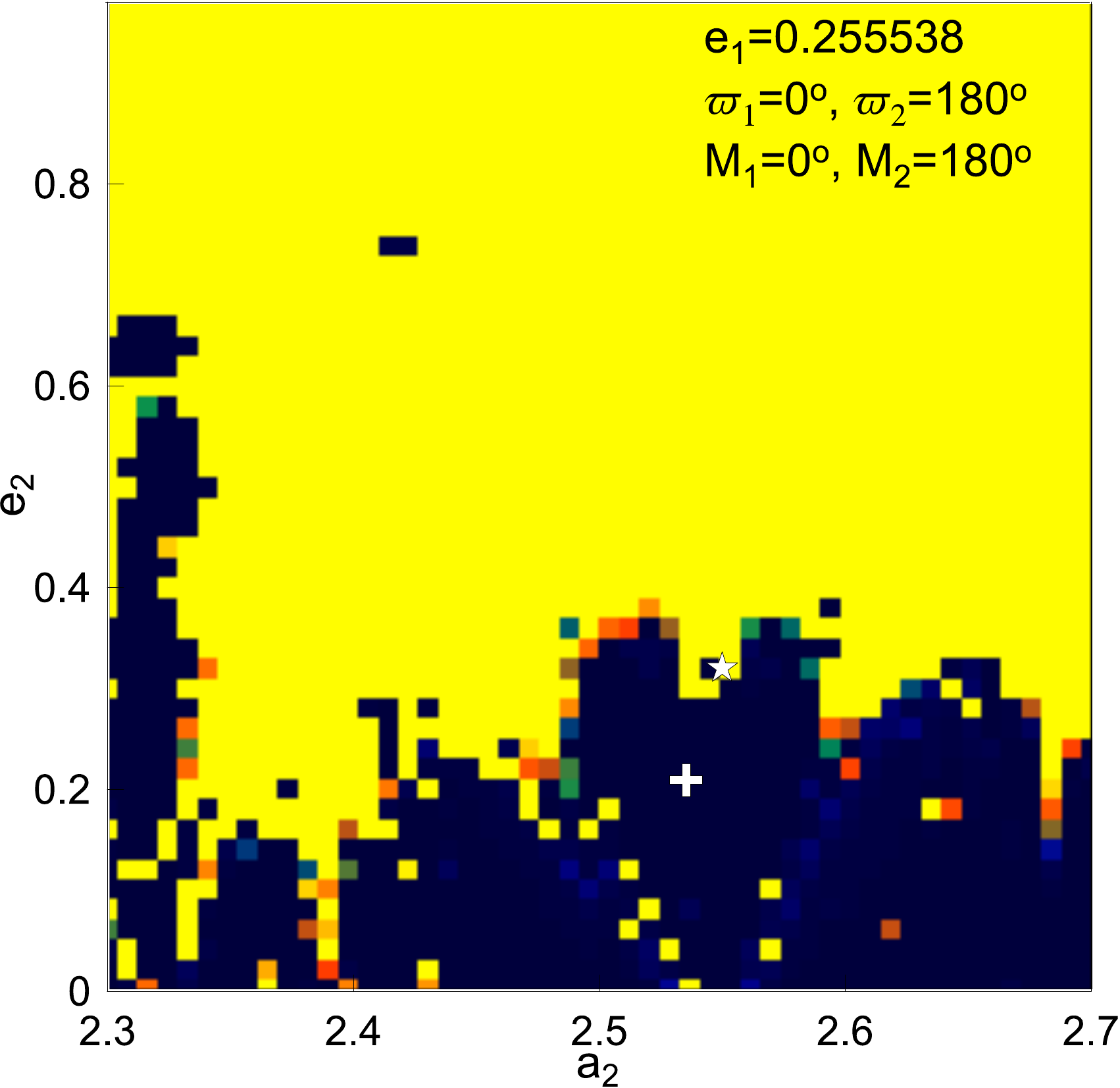}  \\
\textnormal{(e)}  & \;&\textnormal{(f)} \\
\end{array} $
$\begin{array}{c}
\includegraphics[width=3.5cm,height=.6cm]{bar.pdf}\vspace{-2em}
\end{array} $
\end{center}
\caption{DS-maps on the plane $(a_2,e_2)$ in neighbourhood of stable periodic orbits (white crosses), which create the islands of stability that host {\bf a} HD $82943$, {\bf b} HD $73526$, {\bf c} HD $128311$,  {\bf d} HD $60532$, {\bf e} HD $45364$ and {\bf f} HD $108874$ (white stars). The  periodic orbits were chosen so that $e_1$ equals to the eccentricity value of the periodic orbit where the evolution is centered (compare with Fig. \ref{pos}). The fixed orbital elements, yielded by the suitably selected periodic orbits, are shown on each plot.}
\label{mps}
\end{figure*}

In Fig. \ref{hd108874f}, we observe that the semi-major axes and eccentricities evolve regularly around their initial values (\ref{hd108874ic}), while the resonant angles, $\theta_1$ and $\theta_2$  rotate. However, $\Delta\varpi$ librates around the angle $180^{\circ}$, which indicates a secular resonance. Thus, the orbital elements should be revised, in order for a 4/1 MMR to be dynamically confirmed (see Figs. \ref{pos}f,\ref{mps}f).   

\section{Conclusions}\label{sec15}
We reviewed the basic notions of periodic orbits in the GTBP, their association with MMRs through the repetitive relative configuration of the bodies and showed the way the phase space around them is \textit{built} given their linear stability.

The importance of periodic orbits in the field of celestial mechanics has long been highlighted. Stable periodic orbits are of particular importance in planetary dynamics, since even highly eccentric planets being close to an exact MMR  \cite{av16} can survive close encounters and collisions and not least, they can attract planetary systems during their formation and drive their migration process \cite{vat14}. 

Presently, we exhibited the necessity of their knowledge, during fitting procedure of the observational data regarding exoplanets. Phase space for high dimensional dynamical systems, like mutually inclined planets locked in MMR, can be complicated. Periodic orbits can help unravel it and we can complement or revise published data and not least, propose viable parameters, should the observational methods fail to provide.

\vspace{2em}
This research has been co-financed by the European Union (European Social Fund – ESF) and Greek national funds through the Operational Program ``Education and Lifelong Learning'' of the National Strategic Reference Framework (NSRF) - Research Funding Program: Thales. Investing in knowledge society through the European Social Fund. 

%

\end{document}